\documentclass{article}
\usepackage[T1]{fontenc}
\usepackage{lmodern}
\usepackage{microtype}
\usepackage[margin=1in]{geometry}

\usepackage{graphicx}
\usepackage{subcaption}
\usepackage{amssymb}
\usepackage{nccmath}
\usepackage{appendix}
\usepackage{float}
\usepackage{mathtools}

\usepackage{parskip}
\usepackage{xcolor}

\usepackage{natbib}
\usepackage{apalike}
\usepackage[para]{footmisc}
\usepackage{authblk}
\usepackage{amsbsy}

\usepackage{booktabs}
\newtheorem{proposition}{Proposition}
\newtheorem{corollary}{Corollary}

\def\la{\lambda}
\def\elm{\ell^{\mathrm{max}}}

\usepackage{hyperref}
\hypersetup{colorlinks=true, allcolors=blue}

\title{Leveraged positions on decentralized lending platforms}
\author[1]{Bastien Baude\thanks{\href{mailto:bastien.baude@centralesupelec.fr}{bastien.baude@centralesupelec.fr}}}
\author[2, 3]{Vincent Danos\thanks{\href{mailto:vincent.danos@ens.fr}{vincent.danos@ens.fr}}}
\author[4]{Hamza El Khalloufi\thanks{\href{mailto:hamza.el-khalloufi@univ-paris1.fr}{hamza.el-khalloufi@univ-paris1.fr}}}

\affil[1]{Université Paris-Saclay, CentraleSupélec, 91192 Gif-sur-Yvette, France}
\affil[2]{CNRS, École Normale Supérieure, 45 rue d'Ulm, 75005 Paris, France}
\affil[3]{School of Informatics, University of Edinburgh, Edinburgh EH8 9AB, UK}
\affil[4]{Université Paris 1 Panthéon-Sorbonne, 12 place du Panthéon, 75005 Paris, France}

\begin{document}

\bibliographystyle{apalike}

\maketitle

\begin{abstract}
\noindent
We develop a mathematical framework to optimize leveraged staking (``loopy'') strategies in Decentralized Finance (DeFi), in which a staked asset is supplied as collateral, the underlying is borrowed and re-staked, and the loop can be repeated across multiple lending markets. Exploiting the fact that DeFi borrow rates are deterministic functions of pool utilization, we reduce the multi-market problem to a convex allocation problem and obtain closed-form solutions under three interest-rate models: linear, kinked, and adaptive (Morpho's AdaptiveCurveIRM). The framework incorporates market-specific leverage limits, utilization-dependent borrowing costs, and transaction fees. Backtests on the Ethereum and Base blockchains using the largest Morpho wstETH/WETH markets (Jan.~1--Apr.~1, 2025) show that rebalanced leveraged positions can reach up to 6.2\% APY versus 3.1\% for unleveraged staking, with strong dependence on position size and rebalancing frequency. Our results provide a mathematical basis for transparent, automated DeFi portfolio optimization.\\


\end{abstract}

\section{Introduction}\label{section:introduction}

Decentralized finance (DeFi) has emerged as a transformative paradigm in financial markets, offering unprecedented transparency and programmability. Among the various DeFi strategies, leveraged staking, colloquially known as ``loopy'' staking, has gained significant traction. This strategy involves depositing staked assets (such as wstETH) as collateral in lending protocols, borrowing the underlying asset (such as WETH), re-staking it, and repeating the process to amplify exposure to staking yields. While conceptually straightforward, optimizing such strategies across multiple markets with varying interest rate models, liquidity constraints, and transaction costs presents a complex mathematical challenge.

Unlike traditional finance, DeFi protocols operate with complete transparency: all market states, interest rate models, and transaction histories are publicly available on-chain. This transparency enables rigorous mathematical modeling and optimization that would be impossible in opaque traditional markets. Moreover, DeFi interest rate models are deterministic functions of market utilization, what we term ``white-box'' models, allowing us to compute the precise relationship between position size and borrowing costs. This stands in contrast to traditional finance where interest rates are often negotiated or determined by opaque internal models.

The primary contribution of this work is to propose a solution to the optimal allocation problem across multiple leveraged staking markets. We derive closed-form solutions for three widely-used interest rate models: linear rates, kinked rates (as used by Aave \citep{whitepaper2020aavev1}), and adaptive rates (Morpho's AdaptiveCurveIRM \citep{adaptive2023morpho}). A key methodological insight is that any leveraged position can be decomposed into a maximally-leveraged component and an unleveraged (pure staking) component. This decomposition, combined with the aggregation of unleveraged positions across markets, transforms the original \textit{a priori} non-convex optimization problem over allocations and leverage ratios into a convex problem over allocations alone, enabling closed-form solutions. Our framework accounts for market-specific constraints including maximum loan-to-value ratios, variable borrowing rates that depend on pool utilization, and transaction costs. We provide efficient algorithms for computing optimal allocations and validate our theoretical results through backtesting on real market data from the Ethereum and Base blockchains.

Our work builds upon recent advances in DeFi. \citet{whitepaper2020aavev1} introduced the kinked interest rate model for decentralized lending, while \citet{adaptive2023morpho} developed adaptive rate mechanisms that respond dynamically to market conditions. Related work on DeFi optimization includes portfolio optimization under automated market makers and optimal liquidity provision strategies. However, to our knowledge, this is the first work to provide closed-form solutions for multi-market leveraged staking optimization with rigorous treatment of transaction costs and complete liquidity assumptions.

The remainder of this paper is organized as follows. Section~\ref{section:related_work} reviews related work. Section~\ref{section:mathematical_framework} introduces some preliminary definitions and notations. Section~\ref{optimal} formulates the optimization problem and derives closed-form solutions for linear, kinked, and adaptive rate models. Section~\ref{costs} extends the framework to account for transaction costs. Section~\ref{backtest} presents numerical results from backtesting on the Ethereum and Base blockchains. Section~\ref{discussion} concludes with a discussion of limitations and future directions, including game-theoretic considerations when multiple agents employ similar strategies.

\section{Related work}\label{section:related_work}

The mathematical analysis of DeFi lending protocols has attracted growing attention from both the computer science and finance communities. \citet{bartoletti2021sok} provide a systematic overview of lending pools in decentralized finance, introducing a formal model to characterize user interactions, identify vulnerabilities, and analyze potential attacks. Their work establishes foundational abstractions for understanding how lending protocols operate and interact with users.

From an economic perspective, \citet{prat2023contagion} study the propagation of financial shocks in DeFi lending networks using data from the Compound protocol. By constructing balance sheets of liquidity pools and applying contagion models, they characterize how distress cascades through the interconnected positions of borrowers and lenders. Their findings highlight the systemic risks inherent in highly leveraged DeFi positions.

The design of interest rate models for DeFi lending has been studied by \citet{bertucci2024agents}, who analyze agents' behavior on lending platforms and propose a theoretical framework for developing optimal interest rate models. Their work demonstrates that optimal control models with state constraints can generate interest rate policies similar to those used in popular markets, and they show that Morpho's AdaptiveCurveIRM can be interpreted as a nonlinear PD controller. Our work is complementary: while they focus on the protocol's perspective of designing interest rate curves, we focus on the user's perspective of optimizing allocations given existing rate models.

Most closely related to our work, \citet{alexander2024leveraged} analyze the risks of leveraged staking and restaking strategies. Their empirical analysis documents the prevalence of looping strategies and identifies key risk factors including liquidation cascades and rate volatility. Our work complements their risk analysis by providing an optimization framework that accounts for these factors through the size effect and transaction cost mechanisms.

\section{Lending positions}\label{section:mathematical_framework}

In this preliminary section we fix a few notations pertaining to positions on a single lending market $i$.
In the next one, we deal with lending positions that span several lending markets. 

From the point of view of a borrower, a lending market $i$ can be characterized by:
\begin{itemize}
    \item a maximum loan-to-value, denoted by $\mathrm{maxLTV}_{i} < 1$;
    \item a liquidity state, with $\bar{S}_{i}$ and $\bar{B}_{i}$ denoting the total value (using USD as numeraire) supplied to and borrowed from the market (with $\bar{S}_{i} \ge \bar B_{i} > 0$);
    \item an interest rate model mapping the market's utilization rate $\bar{U}_{i} =\bar{B}_{i}/\bar{S}_{i} \leq 1$ to the instantaneous borrow interest rate $b_{i}(\bar{U}_{i})$.
\end{itemize}

\begin{figure}
    \centering
    \includegraphics[scale = 0.6]{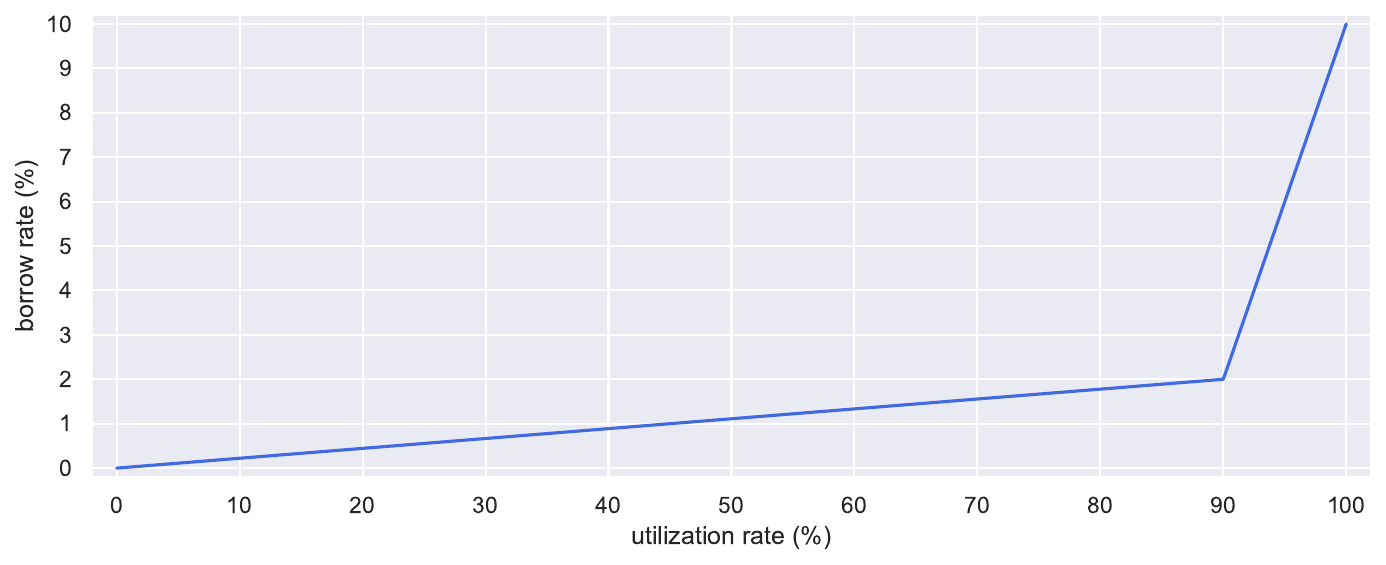}
    \caption{Borrow rate as a function of utilization under the kinked model (illustrative example).}
    \label{fig:kinked_rate_illustration}
\end{figure}
We assume $b_{i}$ is continuous, strictly increasing (higher utilization implies higher borrow rates) and convex. This holds for the interest-rate models considered in this paper: linear and kinked/adaptive models with positive and increasing slopes. Convexity and strict monotonicity ensure that the (separable) objectives considered below are concave---and strictly concave once the budget constraint is used to eliminate the aggregated unleveraged allocation---hence Karush-Kuhn-Tucker (KKT) conditions characterize the unique global optimum and standard one-dimensional root-finders (e.g., Brent's method) behave well. We do not assume differentiability everywhere: kinked/adaptive models are not differentiable at the kink, and first-order conditions there translate into subgradient inequalities. Figure \ref{fig:kinked_rate_illustration} shows an illustrative example of the kinked interest rate model, emphasizing the non-differentiability at the kink.

Suppose we have taken a position $(C_{i}, B_{i})$ on market $i$, that is to say we have deposited a value $C_{i} > 0$ of collateral (in USD), and borrowed a value $B_{i}$ (in USD) of the loanable. Our allocation in market $i$, or the value of our position, denoted by $x_{i}$ (in USD), is defined as:
\begin{equation}\label{eq:single_position}
x_{i} = C_{i} - B_{i} > 0
\end{equation}

The position is subject to the collateralization constraint:
\begin{equation}\label{eq:collat}
B_{i}/C_{i} < \mathrm{maxLTV}_{i}
\end{equation}
If the price of $C_{i}$ relative to $B_{i}$ decreases, and this condition is violated,
the position is liquidated. Because we work with instantaneous USD values (i.e., everything is marked to market at the time of optimization), liquidation dynamics are outside the scope of this static problem; in practice, maintaining a safety buffer is crucial.

It will be convenient to describe such a position by introducing explicitly its leverage $\ell_{i}$ defined as:
\begin{equation}\label{leverage}
\ell_{i} = C_{i}/(C_{i} - B_{i})
\end{equation}
Note that by construction $\ell_{i} \ge 1$. The position $(C_{i}, B_{i})$ can now be described as:
\begin{equation}\label{eq:leveraged_position}
C_{i} = x_{i} \ell_{i}, \quad B_{i} = x_{i} ( \ell_{i} - 1 )
\end{equation}
If $\ell_{i} = 1$, we have borrowed nothing and the entire value of the position is the amount deposited as collateral.

The collateralization constraint above fixes the maximum allowed leverage in market $i$:
\begin{equation}\label{eq:maximum_leverage}
\ell_{i} \leq \frac{1}{1-\mathrm{maxLTV}_{i}}
\end{equation}

The position can be decomposed into:
\begin{itemize}
    \item an unleveraged (staked) sub-position;
    \item a leveraged sub-position with the maximum leverage $\elm_{i} \leq \frac{1}{1-\mathrm{maxLTV}_{i}}$ which we allow internally for a position.
\end{itemize}

Let $x^{1}_{i}$ be the allocation to the leveraged sub-position with maximum leverage and $x^{0}_{i}$ the allocation to the unleveraged (staked) sub-position in market $i$. Since the overall position is the sum of the two sub-positions, we get:
\begin{equation}
C_{i} =  x^{0}_{i}+x^{1}_{i} \elm_{i}, \quad B_{i} = x^{1}_{i} ( \elm_{i} - 1 )
\end{equation}
Solving for $x^{1}_{i}$ and $x^{0}_{i}$ in terms of $x_{i}$ and $\ell_{i}$ yields:
\begin{equation}\label{eq:positions_x_1_x_0}
x^{1}_{i} = x_{i} \frac{\ell_{i} - 1}{\elm_{i} - 1}, \quad 
x^{0}_{i} = x_{i} \frac{\elm_{i} - \ell_i}{\elm_{i} - 1}
\end{equation}

As discussed above, we may choose any internal leverage cap $\elm_{i} \le \frac{1}{1-\mathrm{maxLTV}_{i}}$. In practice we pick $\elm_{i}$ with a safety margin below the theoretical maximum to reduce liquidation risk under adverse price moves of the collateral relative to the loan asset. In the numerical examples, we use Lido's wstETH as collateral. Since wstETH has historically appreciated (almost monotonically) relative to WETH, we set a conservative cap $\elm = 5$, far below the theoretical maximum (typical $\mathrm{maxLTV}$ values for these markets are at least $0.945$, corresponding to a maximum leverage above $18$). With $\elm = 5$ the collateral-to-debt ratio is $C/B = 5/4$ in value, which has historically been ample protection against wstETH depegs.

\section{Optimal capital allocation}\label{optimal}

Let $\xi$ denote our total budget (in USD). Our objective is to allocate $\xi$ across $n$ distinct markets to maximize overall cash flow. We assume that the collateral asset is the same across markets, so all collateral earns the same instantaneous staking rate $s$ (treated as exogenous). The canonical example is wstETH, which accrues a staking yield of approximately $3 \%$ for holders.

For each market $i$, we set an $\elm_{i}$ as explained above. Our decision variables are then the allocation $x_{i}$ (in USD) and the leverage multiplier $\ell_{i}$. We seek to determine the optimal pair $( x_{i}, \ell_{i} )$ for each market $i$ such that the cash flow is maximized, subject to the total budget constraint and any market-specific restrictions (e.g., maximum leverage constraints). The basic optimization problem is as follows:
\begin{equation}\label{eq:problem_x_l}
\begin{aligned}
(x, \ell )^{\ast} = \underset{(x, \ell)}{\mathrm{argmax}} \quad & \sum^{n}_{i=1} x_{i} \ell_{i} s - \sum^{n}_{i=1} x_{i} ( \ell_{i} - 1 ) \hat b_{i} \big ( x_{i} ( \ell_{i} - 1 ) \big ) \\
\textrm{s.t.} \quad & \sum^{n}_{i=1} x_{i} = \xi \\
                    & x_{i} \geq 0, \quad i=1,\ldots,n \\
                    & x_{i}(\ell_{i}-1) \le \bar{S}_{i} - \bar{B}_{i}, \quad i=1,\ldots,n \\
                    & 1 \leq \ell_{i} \leq \elm_{i}, \quad i=1,\ldots,n
\end{aligned}
\end{equation}
with:
\[ \hat{b}_{i}(x) = b_{i} \big ((\bar{B}_{i}+x)/\bar{S}_{i} \big ) \]
that is to say $\hat{b}_{i}(x)$ is the new borrow rate induced by borrowing $x$ on top of the amount $\bar{B}_{i}$ already borrowed. Henceforth, we will write simply $b_i$ for $\hat b_i$ to keep notations simple.

For completeness, we include the feasibility constraint $x_{i}(\ell_{i}-1) \le \bar{S}_{i} - \bar{B}_{i}$, which ensures that additional borrowing does not exceed available liquidity. In practice, borrow rates typically explode as utilization approaches full capacity (see Figure \ref{fig:kinked_rate_illustration}), so this constraint rarely binds; henceforth we will often leave it implicit.

In the objective, the first term represents the staking yield accrued on the collateral, while the second term represents the interest paid on the borrowed amount. The optimization problem is non-trivial because borrowing more increases utilization and therefore increases the marginal borrowing rate $b_{i}$.

If we set $m_{i} = \ell_{i} - 1$, we see that the objective function can be written as:
\begin{equation}
\xi s + \sum^{n}_{i=1} m_{i} x_{i} \times ( \underbrace{s - b_{i}(m_{i} x_{i})}_{\text{carry}} )
\end{equation}
The term $\xi s$ corresponds to the yield of the position where we borrow nothing ($\ell_{i} = 1$, or $m_{i} = 0$). The expression $s - b_{i}(m_{i} x_{i})$ is sometimes called the \emph{carry}. We see that building a looped position on a market is only profitable if the collateral's staking rate exceeds the rate at which one borrows the loanable, that is if the carry is positive.

Problem (\ref{eq:problem_x_l}) does not seem convex in the first instance. However, using (\ref{eq:positions_x_1_x_0}), we can replace $(x_{i}, \ell_{i})$ with $(x^{0}_{i}, x^{1}_{i})$ and obtain an equivalent and clearly convex problem:
\begin{equation}\label{eq:problem_x_1_x_0}
\begin{aligned}
(x^{1}, x^{0})^{\ast} = \underset{(x^{1}, x^{0})}{\mathrm{argmax}} \quad & \sum^{n}_{i=1} x^{0}_{i} s + \sum^{n}_{i=1}  x^{1}_{i} \elm_{i} s - \sum^{n}_{i=1} x^{1}_{i} ( \elm_{i} - 1 ) b_{i} \big ( x^{1}_{i} ( \elm_{i} - 1 ) \big ) \\
\textrm{s.t.} \quad & \sum^{n}_{i=1} x^{1}_{i} + \sum^{n}_{i=1} x^{0}_{i} = \xi \\
                    & x^{1}_{i} \geq 0, \quad i=1,\ldots,n \\
                    & x^{0}_{i} \geq 0, \quad i=1,\ldots,n
\end{aligned}
\end{equation}

(See Appendix~\ref{appendix:equivalence} for details about the equivalence between the two problems.)

The reformulated problem \eqref{eq:problem_x_1_x_0} is a convex optimization problem: the constraints are linear, and the nonlinear objective terms are separable and of the form $-x\,h(x)$ with $h$ strictly increasing and convex, hence strictly concave in $x$.

Since the problem \eqref{eq:problem_x_1_x_0} depends on the sum of the unleveraged allocations, we define:
\begin{equation}\label{eq:sum_unleveraged_positions}
x_{0} = \sum^{n}_{i=1} x^{0}_{i}
\end{equation}
The aggregation of unleveraged allocations allows us to consider the overall unleveraged allocation as a single decision variable. Consequently, our allocation problem involves distributing the total budget across $n$ markets at (a conventional) maximum leverage and one aggregated unleveraged position $x_{0}$. Therefore, the total budget is allocated over $n+1$ positions with the objective of maximizing cash flow. For notational simplicity, we omit the superscript notation and write $x_{i}$ instead of $x^{1}_{i}$ for $i=1,\ldots,n$. The optimization problem \eqref{eq:problem_x_1_x_0} becomes:
\begin{equation}\label{eq:problem_x}
\begin{aligned}
x^{\ast} = \underset{x}{\mathrm{argmax}} \quad 
& x_{0} s + \sum^{n}_{i=1} x_{i} \elm_{i} s - \sum^{n}_{i=1} x_{i} ( \elm_{i} - 1 ) b_{i} \big ( x_{i} ( \elm_{i} - 1 ) \big ) \\
\textrm{s.t.} \quad & x_0+\sum^{n}_{i=1} x_{i} = \xi \\
                    & x_{i} \geq 0, \quad i=0,\ldots,n
\end{aligned}
\end{equation}

Thus reformulated, the problem is in the class of convex allocation problems \citep{patriksson2008survey}. We can solve it using a Lagrange multiplier $\la$ for the equality constraint $x_{0} + \sum^{n}_{i=1} x_{i} = \xi$:
\begin{equation}\label{eq:problem_Lagrange}
\begin{aligned}
x^{\ast}(\la) = \underset{x}{\mathrm{argmax}} \quad & x_{0} (s - \la) + \sum^{n}_{i=1} x_{i} \elm_{i} s - \sum^{n}_{i=1} x_{i} ( \elm_{i} - 1 ) b_{i} \big ( x_{i} ( \elm_{i} - 1 ) \big ) + \la \big ( \xi - \sum^{n}_{i=1} x_{i} \big ) \\
\textrm{s.t.} \quad & x_{i} \geq 0, \quad i=0,\ldots,n
\end{aligned}
\end{equation}
The partial derivative with respect to $x_{0}$ is equal to $s - \la$. Consequently, $\la^{\ast} = s$ if $x_{0} > 0$. Otherwise, we should have $\la^{\ast} > s$. We now examine these two cases separately.

\paragraph{Unsaturated markets with fully leveraged positions \normalfont ($\la^{\ast} > s$)} In this regime, the total budget is not sufficient to saturate\footnote{meaning, had we more budget it would be advantageous to borrow more} all markets, so the unleveraged allocation vanishes ($x^{\ast}_{0} = 0$). Moreover, for each market $i$, the optimal allocation $x^{\ast}_{i}$ is given by the First Order Condition (FOC) obtained from differentiating the objective function in \eqref{eq:problem_Lagrange} with respect to $x_{i}$:
\begin{equation}\label{eq:first_order_condition_i_unsaturated}
\elm_{i} s - (\elm_{i} - 1) \big ( b_{i} \big ( x^{\ast}_{i} ( \elm_{i} - 1 ) \big ) + x^{\ast}_{i} ( \elm_{i} - 1 ) b'_{i} ( x^{\ast}_{i} ( \elm_{i} - 1 ) ) \big ) = \la
\end{equation}
And the optimal Lagrange multiplier $\la^{\ast}$ is the solution to the budget constraint:
\begin{equation}\label{eq:budget_constraint_unsaturated}
\sum^{n}_{i=1} x^{\ast}_{i}(\la^{\ast}) = \xi
\end{equation}

\paragraph{Saturated markets with strictly positive unleveraged position \normalfont ($\la^{\ast} = s$)} For each market $i$, we determine the optimal allocation $x^{\ast}_{i}$ by differentiating the objective function in \eqref{eq:problem_Lagrange} with respect to $x_{i}$ and imposing the FOC:
\begin{equation}\label{eq:first_order_condition_i_saturated}
b_{i} \big ( x^{\ast}_{i} ( \elm_{i} - 1 ) \big ) + x^{\ast}_{i} ( \elm_{i} - 1 ) b'_{i} \big ( x^{\ast}_{i} ( \elm_{i} - 1 ) \big ) = s
\end{equation}
The unleveraged allocation $x^{\ast}_{0}$ follows from the budget constraint:
\begin{equation}\label{eq:x_0_saturated}
x^{\ast}_{0} = \xi - \sum^{n}_{i=1} x^{\ast}_{i}
\end{equation}
In this regime, all markets are saturated, and the residual budget is allocated to the unleveraged position.

Since \eqref{eq:first_order_condition_i_saturated} is recovered from \eqref{eq:first_order_condition_i_unsaturated} with $\la = s$, henceforth we focus exclusively on \eqref{eq:first_order_condition_i_unsaturated}.

Above we have used $b'_{i}$ as if the rate function was everywhere differentiable. Let $B_{i} = (\elm_{i} - 1) x_{i}$ (the amount we borrow) and  define $g_{i}(B_{i}) = B_i\,b_{i}(B_{i})$. When $b_{i}$ is kinked, $g_{i}$ may not be differentiable at $B_{i}^{\mathrm{kink}} = \bar{S}_{i} u^{\ast} - \bar{B}_{i}$.
The KKT condition is now a subgradient one:
\begin{equation}
\la \in \elm_{i} s-(\elm_{i} - 1)\,\partial g_{i}(B_{i})
\end{equation}
so if $\la$ lies between the left and right marginal costs at $B_{i}^{\mathrm{kink}}$, the maximizer is attained at the kink boundary $B_{i} = B_{i}^{\mathrm{kink}}$, i.e. $x_{i} = B_{i}^{\mathrm{kink}}/(\elm_{i} - 1)$. This will show clearly in our solution of the kinked rate model below (see Section \ref{kinked_section}).

\subsection{Algorithm}\label{algorithm}

First, we compute the optimal allocations from \eqref{eq:first_order_condition_i_unsaturated} with $\la = s$, denoted by $x^{\ast, (0)}$. If the budget constraint holds, that is to say if $\sum^{n}_{i=1} x^{\ast, (0)}_{i} \leq \xi$, the solution is valid, and the unleveraged allocation follows from: $x^{\ast, (0)}_{0} = \xi - \sum^{n}_{i=1} x^{\ast, (0)}_{i}$.

Else, we compute the optimal allocations from \eqref{eq:first_order_condition_i_unsaturated}, denoted by $x^{\ast, (1)}(\la)$ given $\la$, for which $x^{\ast, (1)}_{0} = 0$. The optimal Lagrange multiplier $\la^{\ast}$ is determined from the budget constraint:
\begin{equation}\label{eq:f}
\sum^{n}_{i=1} x^{\ast,(1)}_{i}(\la^{\ast}) = \xi
\end{equation}
Closed-form solutions may exist depending on the interest rate model; otherwise, $\la^{\ast}$ is determined numerically, e.g., using Brent’s algorithm.

\subsection{Interpreting $\la^{\ast}$}

One can think of $\la^{\ast}$, the optimal Lagrange multiplier, as a way to make the leveraged strategy artificially and gradually less profitable, up until the point where all markets become saturated under the budget constraint. Suppose that the same maximum allowed leverage $\elm$ is applied to all markets. Then, (\ref{eq:first_order_condition_i_unsaturated}) becomes equivalent to (\ref{eq:first_order_condition_i_saturated}) using a modified staking rate:
\begin{equation}\label{eq:effective_staking_rate} 
s^{a}(\la) = s + \frac{s - \la}{\ell^{\text{max}} - 1} 
\end{equation}

Since $\la^{\ast} \geq s$, it follows that $s^{a}(\la^{\ast}) \leq s$. To find $\la^{\ast}$, the algorithm starts from $\la = s$ and increases it, which is equivalent to decreasing $s^{a}$, until the budget constraint is satisfied. Once it is found, the optimal strategy derived from (\ref{eq:first_order_condition_i_unsaturated}) (unsaturated markets) is equivalent to the one obtained from (\ref{eq:first_order_condition_i_saturated}) (all markets saturated) under the artificial staking rate $s^{a}(\la^{\ast})$ instead of the original $s$.

A similar interpretation applies on the borrow side. Increasing the Lagrange multiplier can be seen as introducing artificial (and higher) borrowing rates, which are gradually raised until all markets are saturated. This interpretation, unlike its counterpart on the staking side, holds even when markets have different leverage limits.

\subsection{Linear rate}\label{linear_section}

Using the notations introduced in \citet{whitepaper2020aavev1}, the linear model reads:
\begin{equation}\label{eq:linear_rate_model}
b_{i} (B_{i}) = r_{base} + \frac{\bar{B}_{i} + B_{i}}{\bar{S}_i u^{*}} r_{slope1}
\end{equation}
where $u^{*}\in(0,1)$ is called the target utilization, 
$r_{base} \geq 0$ and $r_{slope1} > 0$. At target utilization, the borrow rate is $r_{base} + r_{slope1}$.

With a linear rate model, our problem becomes a so-called ``water-filling'' problem \cite[Example 5.2]{boyd}. We now state its general closed-form solution.
\medskip
\begin{proposition}[Linear rate]\label{pr:linear_rate_solution} Under the linear rate model \eqref{eq:linear_rate_model}, the optimal solution to \eqref{eq:first_order_condition_i_unsaturated} is given by:
\begin{equation}\label{eq:linear_rate_solution}
x^{*}_{i}(\la) = \alpha_{i} \big [ \beta_{i} - \la \big ]^{+}
\end{equation}
where,
\begin{equation}\label{eq:parameters_linear_rate_solution}
\alpha_{i} = \frac{\bar{S}_{i} u^{*}}{2 r_{slope1} ( \elm_{i} - 1 )^{2}}, \quad \beta_{i} = \elm_{i} s - ( \elm_{i} - 1 ) \big ( r_{base} + \frac{\bar{B}_{i}}{\bar{S}_{i} u^{*}} r_{slope1} \big )
\end{equation}
for $i = 1, \ldots, n$ and the optimal Lagrange multiplier reads:
\begin{equation}\label{eq:optimal_lambda}
\la^{*} = \frac{\sum^{k}_{j=1} \alpha_{j} \beta_{j} - \xi}{\sum^{k}_{j=1} \alpha_{j}}
\end{equation}
Without loss of generality, we assume that the markets are ordered such that:
\begin{equation}\label{eq:beta_sorted}
\beta_{1} \geq \beta_{2} \geq \cdots \geq \beta_{n}
\end{equation}
and the index $k \in \{ 1, \ldots, n \}$ is determined by the conditions (with by convention $\varphi_{n+1} = +\infty$):
\begin{equation}\label{eq:phi_condition}
\varphi_{k} < \xi \leq \varphi_{k+1}
\end{equation}
where,
\begin{equation}
\varphi_{k} = \sum^{k}_{j=1} \alpha_{j} \big [ \beta_{j} - \beta_{k} \big ]
\end{equation}
\end{proposition}
The proof is provided in Appendix \ref{appendix:proof_linear_rate_solution}.

\subsection{Kinked rate}\label{kinked_section}

The kinked model reads:
\begin{equation}\label{eq:kinked_rate_model}
b_{i} (B_{i}) = \left\{
    \begin{array}{ll}
        \displaystyle r_{base} + \frac{\bar{B}_{i} + B_{i}}{\bar{S}_{i} u^{*}} r_{slope1} & \mbox{if } \bar{B}_{i} + B_{i} < \bar{S}_{i} u^{*} \\
        \displaystyle r_{base} + r_{slope1} + \frac{\bar{B}_{i} + B_{i} - \bar{S}_{i} u^{*}}{\bar{S}_{i} (1 - u^{*})} r_{slope2} & \mbox{if } \bar{B}_{i} + B_{i} \geq \bar{S}_{i} u^{*}
    \end{array}
\right.
\end{equation}
where $u^{*}\in(0,1)$, $r_{base} \geq 0$, $r_{slope1} > 0$ and $r_{slope2} > 0$. Again, we use the notation of \citet{whitepaper2020aavev1}. The parameter $r_{slope1}$ is normalized such that the rate at target utilization $u^{*}$ equals $r_{base} + r_{slope1}$, while $r_{slope2}$ is normalized so that the rate at full utilization equals $r_{base} + r_{slope1} + r_{slope2}$.

We now state the general closed-form solution under kinked interest rate models.
\medskip
\begin{proposition}[Kinked rate]\label{pr:kinked_rate_solution} Under the kinked rate model \eqref{eq:kinked_rate_model} and assuming $r_{slope1} < \frac{u^{*}}{1 - u^{*}} r_{slope2}$, if $\bar{S}_{i} u^{*} - \bar{B}_{i} > 0$, i.e., when the current utilization is below the target rate, the optimal solution to \eqref{eq:first_order_condition_i_unsaturated} is given by:
\begin{equation}\label{eq:kinked_rate_solution_below}
x^{*}_{i}(\la) = \left\{
    \begin{array}{ll}
        \displaystyle \alpha^{2}_{i} \big [ \beta^{2}_{i} - \la \big ]^{+} & \mbox{if } \la < \la^{2}_{i} \\
        \displaystyle \frac{\bar{S}_{i} u^{*} - \bar{B}_{i}}{\elm_{i} - 1} & \mbox{if } \la^{2}_{i} \leq \la \leq \la^{1}_{i} \\
        \displaystyle \alpha^{1}_{i} \big [ \beta^{1}_{i} - \la \big ]^{+} & \mbox{if } \la^{1}_{i} < \la
    \end{array}
\right.
\end{equation}
where,
\begin{equation}
\begin{aligned}\label{eq:parameters_kinked_rate_solution}
\alpha^{1}_{i} & = \frac{\bar{S}_{i} u^{*}}{2 r_{slope1} ( \elm_{i} - 1 )^{2}}, \quad \beta^{1}_{i} = \elm_{i} s - ( \elm_{i} - 1 ) \big ( r_{base} + \frac{\bar{B}_{i}}{\bar{S}_{i} u^{*}} r_{slope1} \big ) \\
\alpha^{2}_{i} & = \frac{\bar{S}_{i} (1 - u^{*})}{2 r_{slope2} ( \elm_{i} - 1 )^{2}}, \quad \beta^{2}_{i} = \elm_{i} s - ( \elm_{i} - 1 ) \big ( r_{base} + r_{slope1} + \frac{\bar{B}_{i} - \bar{S}_{i} u^{*}}{\bar{S}_{i} (1 - u^{*})} r_{slope2} \big )
\end{aligned}
\end{equation}
and,
\begin{equation}
\begin{aligned}\label{eq:lambdas_kinked_rate_solution}
\la^{1}_{i} & = \elm_{i} s - ( \elm_{i} - 1 ) \big ( r_{base} + r_{slope1} + (\bar{S}_{i} u^{*} - \bar{B}_{i}) \frac{r_{slope1}}{\bar{S}_{i} u^{*}}  \big ) \\
\la^{2}_{i} & = \elm_{i} s - ( \elm_{i} - 1 ) \big ( r_{base} + r_{slope1} + (\bar{S}_{i} u^{*} - \bar{B}_{i}) \frac{r_{slope2}}{\bar{S}_{i} (1 - u^{*})}  \big )
\end{aligned}
\end{equation}
Otherwise, if $\bar{S}_{i} u^{*} - \bar{B}_{i} \leq 0$, i.e., when the current utilization is at or above the target rate, the optimal solution reads:
\begin{equation}\label{eq:kinked_rate_solution_above}
x^{*}_{i}(\la) = \alpha^{2}_{i} \big [ \beta^{2}_{i} - \la \big ]^{+}
\end{equation}
\end{proposition}
The proof is provided in Appendix \ref{appendix:proof_kinked_rate_solution}.

\subsection{Adaptive rate}\label{adaptive_section}

In contrast to the linear and kinked rate models, which depend solely on the current utilization rate, the Morpho interest rate model \citep{adaptive2023morpho}, called AdaptiveCurveIRM (abbreviated as adaptive rate) also depends on the previous state of the liquidity pool. Specifically, the model is given by $r(u,t) = r^{\mathrm{target}}_{t} \text{curve}(u)$ where,
\begin{equation}
r^{\mathrm{target}}_{t} = r^{\mathrm{target}}_{t_{last}} \text{speed}(t), \quad \text{speed}(t) = e^{k_{p} \text{error}(u_{t_{last}}) (t - t_{last})}, \quad 
\text{error}(u) = \left\{
\begin{array}{ll}
    \frac{u - u^{*}}{u^{*}} & \mbox{if } u < u^{*} \\
    \frac{u - u^{*}}{1 - u^{*}} & \mbox{if } u \geq u^{*}
\end{array}
\right.
\end{equation}
where $k_{p} > 0$. The time $t_{last}$ corresponds to the last interaction with the pool (deposit, withdrawal, borrow, or repay) and $u_{t_{last}}$ denotes the associated utilization rate. In addition, we have:
\begin{equation}
\text{curve}(u) = \left\{
    \begin{array}{ll}
        \big ( 1 - \frac{1}{k_{d}} \big ) \text{error}(u) + 1 & \mbox{if } u < u^{*} \\
        \big ( k_{d} - 1 \big ) \text{error}(u) + 1 & \mbox{if } u \geq u^{*}
    \end{array}
\right.
\end{equation}
where $u^{*} \in (0,1)$ and $k_{d} > 1$.

The adaptive rate model can be reformulated as follows:
\begin{equation}\label{eq:adaptive_rate_model}
b_{i} (B_{i}) = \left\{
    \begin{array}{ll}
        \displaystyle r^{\mathrm{target}}_{t} \big [ 1 + \big ( 1 - \frac{1}{k_{d}} \big ) \frac{\bar{B}_{i} + B_{i} - \bar{S}_{i} u^{*}}{\bar{S}_{i} u^{*}} \big ] & \mbox{if } \bar{B}_{i} + B_{i} < \bar{S}_{i} u^{*} \\
        \displaystyle r^{\mathrm{target}}_{t} \big [ 1 + \big ( k_{d} - 1 \big ) \frac{\bar{B}_{i} + B_{i} - \bar{S}_{i} u^{*}}{\bar{S}_{i} (1 - u^{*} ) } \big ] & \mbox{if } \bar{B}_{i} + B_{i} \geq \bar{S}_{i} u^{*}
    \end{array}
\right.
\end{equation}

By omitting the dynamic feature of the adaptive rate---which is justified in our case since the optimization problem is static in time---the adaptive model is a reparametrization of the kinked model. Nonetheless, we still provide the general closed-form solution under adaptive interest rate models, as we believe it may be useful for practitioners.
\medskip
\begin{corollary}[Adaptive rate]\label{co:adaptive_rate_solution} Under the adaptive rate model \eqref{eq:adaptive_rate_model} and assuming $\frac{1 - u^{*}}{u^{*}} < k_{d}$, if $\bar{S}_{i} u^{*} - \bar{B}_{i} > 0$, i.e., when the current utilization is below the target rate, the optimal solution to \eqref{eq:first_order_condition_i_unsaturated} is given by:
\begin{equation}\label{eq:adaptive_rate_solution_below}
x^{*}_{i}(\la) = \left\{
    \begin{array}{ll}
        \displaystyle \alpha^{2}_{i} \big [ \beta^{2}_{i} - \la \big ]^{+} & \mbox{if } \la < \la^{2}_{i} \\
        \displaystyle \frac{\bar{S}_{i} u^{*} - \bar{B}_{i}}{\elm_{i} - 1} & \mbox{if } \la^{2}_{i} \leq \la \leq \la^{1}_{i} \\
        \displaystyle \alpha^{1}_{i} \big [ \beta^{1}_{i} - \la \big ]^{+} & \mbox{if } \la^{1}_{i} < \la
    \end{array}
\right.
\end{equation}
where,
\begin{equation}
\begin{aligned}\label{eq:parameters_adaptive_rate_solution}
\alpha^{1}_{i} & = \frac{\bar{S}_{i} u^{*}}{2 r^{\mathrm{target}}_{t} (1 - \frac{1}{k_{d}}) ( \elm_{i} - 1 )^{2}}, \quad \beta^{1}_{i} = \elm_{i} s - ( \elm_{i} - 1 ) r^{\mathrm{target}}_{t} \big ( \frac{1}{k_{d}} + \frac{\bar{B}_{i}}{\bar{S}_{i} u^{*}} (1 - \frac{1}{k_{d}}) \big ) \\
\alpha^{2}_{i} & = \frac{\bar{S}_{i} (1 - u^{*})}{2 r^{\mathrm{target}}_{t} (k_{d} - 1) ( \elm_{i} - 1 )^{2}}, \quad \beta^{2}_{i} = \elm_{i} s - ( \elm_{i} - 1 ) r^{\mathrm{target}}_{t} \big ( 1 + \frac{\bar{B}_{i} - \bar{S}_{i} u^{*}}{\bar{S}_{i} (1 - u^{*})} (k_{d} - 1) \big )
\end{aligned}
\end{equation}
and,
\begin{equation}\label{eq:lambdas_adaptive_rate_solution}
\begin{aligned}
\la^{1}_{i} & = \elm_{i} s - ( \elm_{i} - 1 ) r^{\mathrm{target}}_{t} \big ( 1 + (\bar{S}_{i} u^{*} - \bar{B}_{i}) \frac{(1 - \frac{1}{k_{d}})}{\bar{S}_{i} u^{*}}  \big ) \\
\la^{2}_{i} & = \elm_{i} s - ( \elm_{i} - 1 ) r^{\mathrm{target}}_{t} \big ( 1 + (\bar{S}_{i} u^{*} - \bar{B}_{i}) \frac{(k_{d} - 1)}{\bar{S}_{i} (1 - u^{*})}  \big )
\end{aligned}
\end{equation}
Otherwise, if $\bar{S}_{i} u^{*} - \bar{B}_{i} \leq 0$, i.e., when the current utilization is at or above the target rate, the optimal solution reads:
\begin{equation}\label{eq:adaptive_rate_solution_above}
x^{*}_{i}(\la) = \alpha^{2}_{i} \big [ \beta^{2}_{i} - \la \big ]^{+}
\end{equation}
\end{corollary}
The proof is provided in Appendix~\ref{appendix:proof_adaptive_rate_solution}. In practice, the values $u^{*} = 0.9$ and $k_{d} = 4$ are hardcoded into the smart contract, and they satisfy the condition $\frac{1 - u^{*}}{u^{*}} < k_{d}$.

\section{Dealing with transaction fees}\label{costs}

\subsection{General approach}

Suppose we already hold a position $\bar{x}$, and we wish to move to a new position $x$. Because of the cost of moving capital around, it may not be profitable to move even if $x$ has a higher yield. Let us write $c(x, \bar{x})$ for the cost of rebalancing the portfolio from the current one to $x$. Let $T$ be our investment horizon, that is to say the typical duration we expect to hold a position.

Our basic optimization problem \eqref{eq:problem_x} can be modified to account for transaction fees:
\begin{equation}\label{eq:problem_x_fees}
\begin{aligned}
x^{\ast} = \underset{x}{\mathrm{argmax}} \quad 
& 
\left\{
- \frac{1}{T} c(x, \bar{x}) +
\left(
x_{0} - \bar{x}_{0}
+ \sum^{n}_{i=1} (x_{i} - \bar{x}_{i}) \elm_{i}\right) s
\right.
\\
&
\left. 
- \sum^{n}_{i=1}  (\elm_{i} - 1) \Big [ 
x_{i} b_{i} \big(x_{i}(\elm_{i} - 1)\big) 
- \bar{x}_{i} b_{i} \big(\bar{x}_{i}(\elm_{i} - 1)\big) 
%- \bar{x}_{i} b_{i} ( 0 ) 
\Big ]
\right\}
\\
\textrm{s.t.} \quad 
& x_{0} + \sum^{n}_{i=1} x_{i} = \xi \\
& x_{i} \geq 0, \quad i=0,\ldots,n
\end{aligned}
\end{equation}
In words, we seek the position $x$ that best trades-off expected yield improvements against rebalancing costs. This trade-off depends on the choice of $T$. The shorter $T$ is, the smaller the optimal rebalancing will be. In addition, $T$ should be small, as compounding effects are neglected in \eqref{eq:problem_x_fees}. (In practice, a natural choice for $T$ is the time it takes for the disparity of borrow rates to dissipate.)

We choose to model transaction fees as follows:
\begin{equation}\label{eq:transaction_cost}
c(x, \bar{x}) = 
\gamma(x, \bar{x})\,
\Big | x_{0} - \bar{x}_{0} + \sum^{n}_{i=1} (x_{i}-\bar{x}_{i}) \elm_{i}  
%- \sum^{n}_{i=1}  \elm_{i} 
\Big |
\end{equation}
and,
\begin{equation}\label{eq:fee_rate}
\gamma(x, \bar{x}) = \left\{
        \begin{array}{ll}
            \displaystyle \gamma^{+} & \mbox{if } \bar{x}_{0} + \sum^{n}_{i=1} \bar{x}_{i} \elm_{i} < x_{0} + \sum^{n}_{i=1} x_{i} \elm_{i} \\
            \displaystyle \gamma^{-} & \mbox{if } \bar{x}_{0} + \sum^{n}_{i=1} \bar{x}_{i} \elm_{i} \geq x_{0} + \sum^{n}_{i=1} x_{i} \elm_{i}
        \end{array}
    \right.
\end{equation}

Note that our simple notion of cost only depends on the change in total collateral amount; moving collateral across markets is assumed to be free, which isn’t true operationally, because of so-called (zero-order) ``gas costs''. We distinguish between the cases where the total collateral increases or decreases. In the former case we use $\gamma^{+}$; in the latter we use $\gamma^{-}$.

To see why the distinction is worth making in practice we can take the example of Lido's wstETH as collateral. To acquire it starting from WETH, one can convert WETH directly without fees. Not so in the other direction where Lido redemption protocol forces a variable delay and selling wstETH for WETH can be done at a small discount on exit markets. Thus in this case there is a strong asymmetry between the two directions as $\gamma^{+} = 0$, while $\gamma^{-} > 0$ is typically of the order of a couple of basis points (bps).

Our transaction fee model is proportional to the change in total collateral, with constant rates $\gamma^{+}$ and $\gamma^{-}$. This formulation implicitly neglects slippage on exit markets. To account for linear slippage—the simplest extension—rates $\gamma^{+}$ and $\gamma^{-}$ should be linear in the change in total collateral. However, this leads to a non-separable objective function, which precludes closed-form solutions.

\subsection{Algorithm}

To solve the problem, we first assume that the total amount of collateral increases. That is to say: $\bar{x}_{0} + \sum^{n}_{i=1} \bar{x}_{i} \elm_{i} < x_{0} + \sum^{n}_{i=1} x_{i} \elm_{i}$. This means that:
\begin{equation}
\frac{1}{T}\, c(x, \bar{x}) = \frac{\gamma^{+}}{T}\left(x_{0} - \bar{x}_{0} + \sum^{n}_{i=1} (x_{i}-\bar{x}_{i}) \elm_{i}\right) 
\end{equation}
Hence, we are back in the case of Section \ref{optimal}, except that we use an effective staking rate:
\begin{equation}
s^{+} = s - \frac{\gamma^{+}}{T}
\end{equation}
If the resulting allocation satisfies the initial inequality, the solution is valid. Else, we compute the optimal allocation with $s^{-} = s + \frac{\gamma^{-}}{T}$. If neither solution is valid, then the optimal allocation is $\bar{x}$, i.e., we should not change our position.

\section{Numerical results}\label{backtest}

\subsection{Data processing}

\begin{figure}
    \centering
    \includegraphics[scale = 0.6]{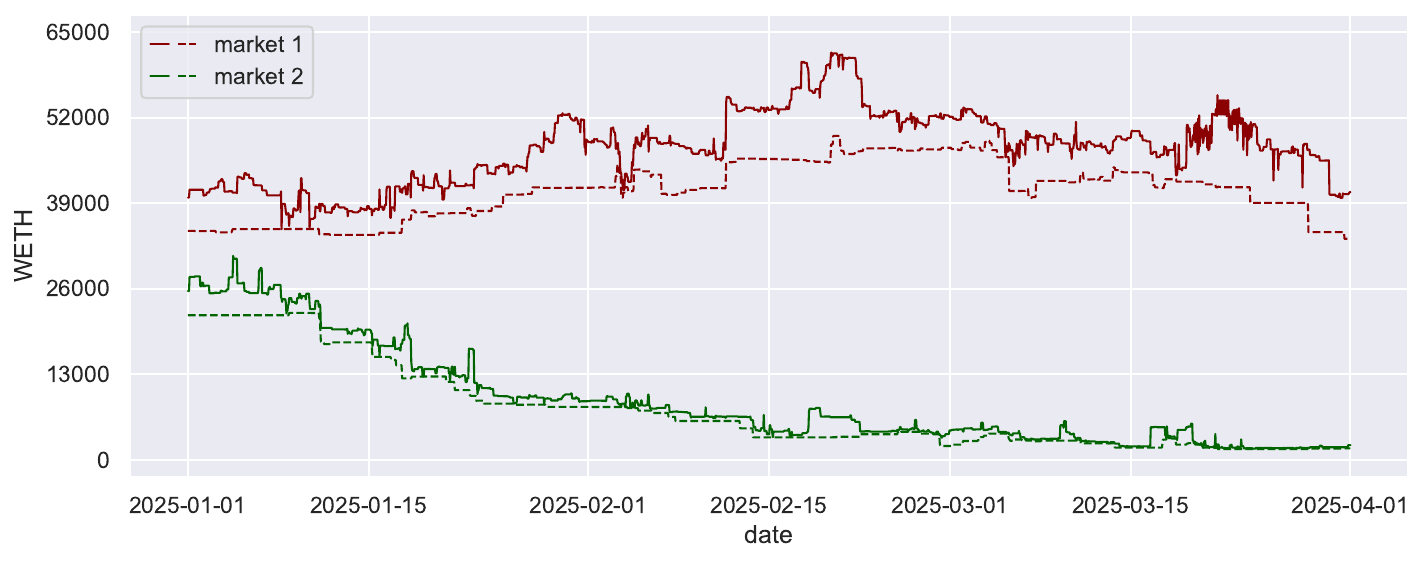}
    \caption{Evolution of WETH reserves (solid line: supplied funds; dashed line: borrowed fund) for the two largest wstETH/WETH markets on Morpho on the Ethereum blockchain from January 1, 2025 to April 1, 2025.}
    \label{fig:evolution_ethereum_market_reserves}
\end{figure}

\begin{table}
\centering
\begin{tabular}{cccc}
\toprule
market & ID & creation date & LLTV ($\%$) \\
\midrule
1 & \texttt{6becf9b4-3c85-40bf-9938-196812e034a3} & March 14, 2024 & $96.5$ \\
2 & \texttt{928c009a-d217-42f7-9d3a-45bb6c8d71f9} & March 25, 2024  & $94.5$ \\
\bottomrule
\end{tabular}
\caption{IDs, creation dates and LLTVs of the two largest wstETH/WETH markets on Morpho on the Ethereum blockchain.}
\label{tab:ethereum_markets}
\end{table}
We backtest the ``loopy'' strategy on real market data from the Ethereum blockchain. We first describe the dataset and preprocessing steps before turning to the results. We focus on the largest Morpho markets for the wstETH/WETH pair over the period from January 1, 2025, to April 1, 2025. The data is retrieved from Morpho's GraphQL service.\footnote{\url{https://api.morpho.org/graphql}} We select the two largest markets in terms of supplied assets; their characteristics are summarized in Table \ref{tab:ethereum_markets}. The third-largest market is negligible in comparison and therefore excluded from the analysis. Figure \ref{fig:evolution_ethereum_market_reserves} shows the evolution of supplied and borrowed WETH in the selected markets. At the beginning of the period, both markets had comparable levels of supplied assets. However, the second market declined substantially over time, and by April 1, 2025, the first market had become nearly $20$ times larger. This difference is likely due to the higher maximum LTV in the first market, which allows for greater leverage capacity (see Table \ref{tab:ethereum_markets}).

\begin{figure}
    \centering
    \begin{subfigure}{\linewidth} 
        \centering
        \includegraphics[scale = 0.6]{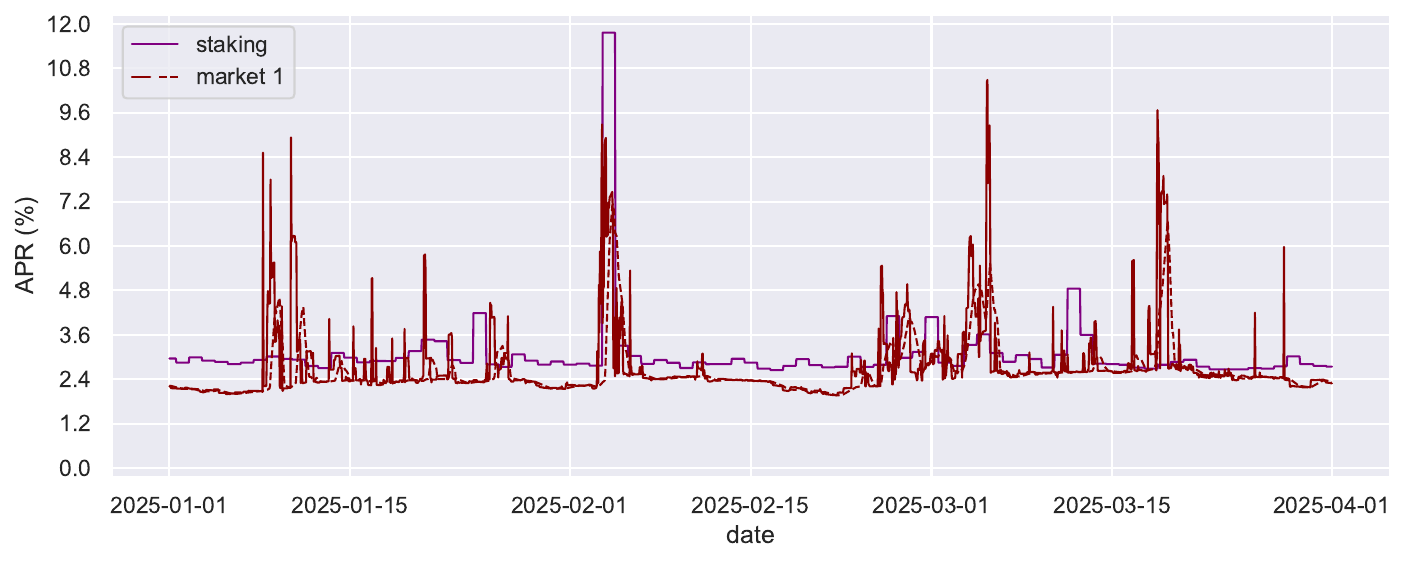}
        \caption{market $1$}
    \end{subfigure}
    \vfill
    \begin{subfigure}{\linewidth}
        \centering
        \includegraphics[scale = 0.6]{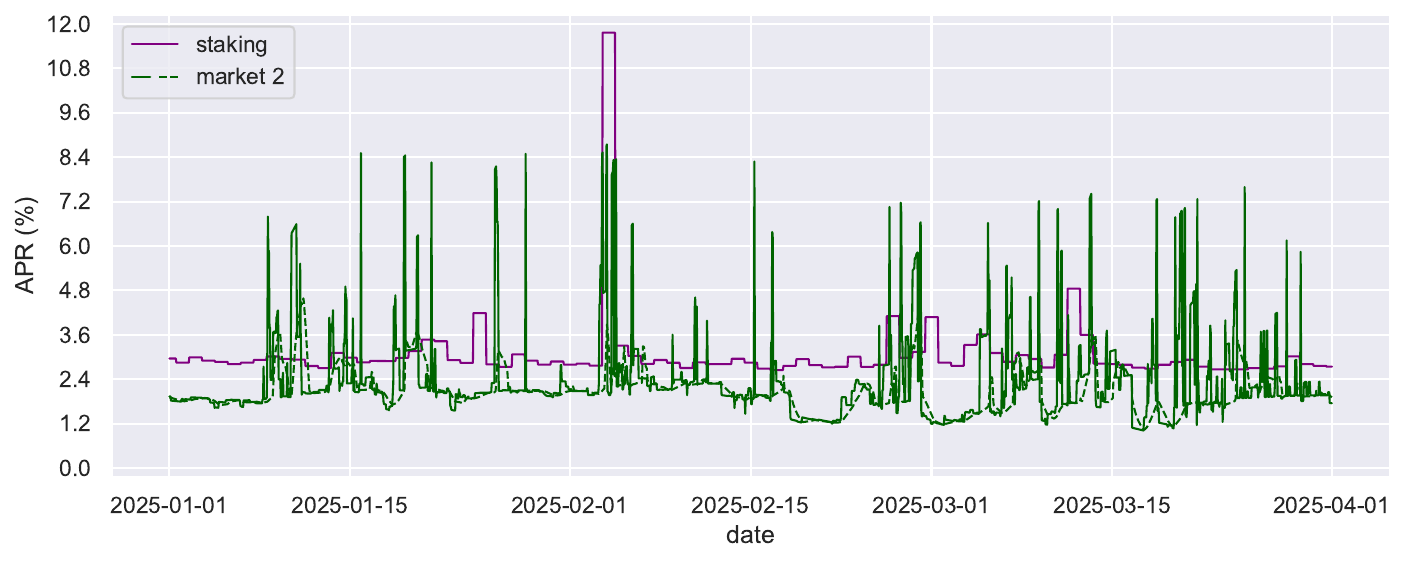}
        \caption{market $2$}
    \end{subfigure}
    \caption{Evolution of the interest rate (solid line: effective rate; dashed line: rate at target) for the two largest wstETH/WETH markets on Morpho on the Ethereum blockchain, compared to the staking rate from January 1, 2025 to April 1, 2025.}
    \label{fig:evolution_ethereum_market_rates}    
\end{figure}

Figure \ref{fig:evolution_ethereum_market_rates} shows the evolution of the borrowing rate and the rate at target for both markets on an hourly basis, compared to the Lido staking rate over the same period. The staking rate data is retrieved via Lido's The Graph service\footnote{\url{https://github.com/lidofinance/lido-subgraph}} on a daily basis. We can identify periods where the borrowing rate is lower than the staking rate, suggesting that a ``loopy’’ strategy would have been profitable, as well as highly volatile periods where this is no longer the case. To reduce noise in the backtest, we apply a one-day moving average to the borrowing rate. In contrast, the staking rate, which is less volatile, is used as-is.

\subsection{Backtesting}

We perform a backtest of the ``loopy’’ strategy using the dataset described above. Two budget configurations are considered: a low budget of \$10k, which has a moderate impact on the liquidity pools, and a high budget of \$10m, whose impact is significant. In addition, the backtest is also conducted with both hourly and daily rebalancing to evaluate the effect of the rebalancing frequency on the strategy’s performance. Table \ref{tab:ethereum_strategies} reports the resulting APYs across the different configurations over the backtesting period.

\begin{table}
\centering
\begin{tabular}{ccccc}
\toprule
strategy & initial investment ($\$$) & rebalancing frequency & $\elm$ & APY ($\%$) \\
\midrule
loopy (low cap, 1h-freq) & $10 \text{k}$ & $1 \text{h}$ & $5$ & $6.2$ \\
loopy (low cap, 1d-freq) & $10 \text{k}$ & $1 \text{d}$ & $5$ & $5.8$ \\
loopy (high cap, 1h-freq) & $10 \text{m}$ & $1 \text{h}$ & $5$ & $3.7$ \\
loopy (high cap, 1d-freq) & $10 \text{m}$ & $1 \text{d}$ & $5$ & $3.7$ \\
staking & $\cdot$ & $\cdot$ & $1$ & $3.1$ \\
\bottomrule
\end{tabular}
\caption{Characteristics and performance of the leveraged strategy compared to staking on the Ethereum blockchain from January 1, 2025 to April 1, 2025.}
\label{tab:ethereum_strategies}
\end{table}
First, we assume a zero-fee setting in the backtest. While this assumption is not realistic in practice, it provides a benchmark for evaluating the ideal performance of the strategy in the absence of transaction costs.
\begin{figure}
    \centering
    \begin{subfigure}{\linewidth} 
        \centering
        \includegraphics[scale = 0.6]{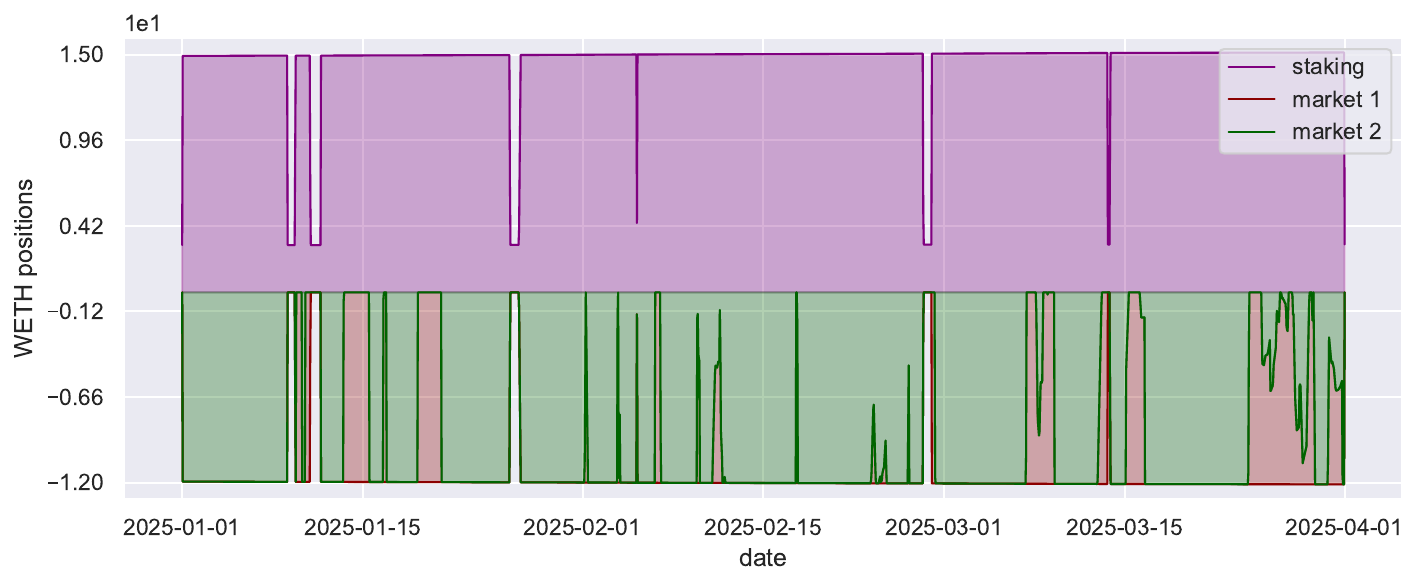}
        \caption{1h-freq rebalancing}
    \end{subfigure}
    \vfill
    \begin{subfigure}{\linewidth}
        \centering
        \includegraphics[scale = 0.6]{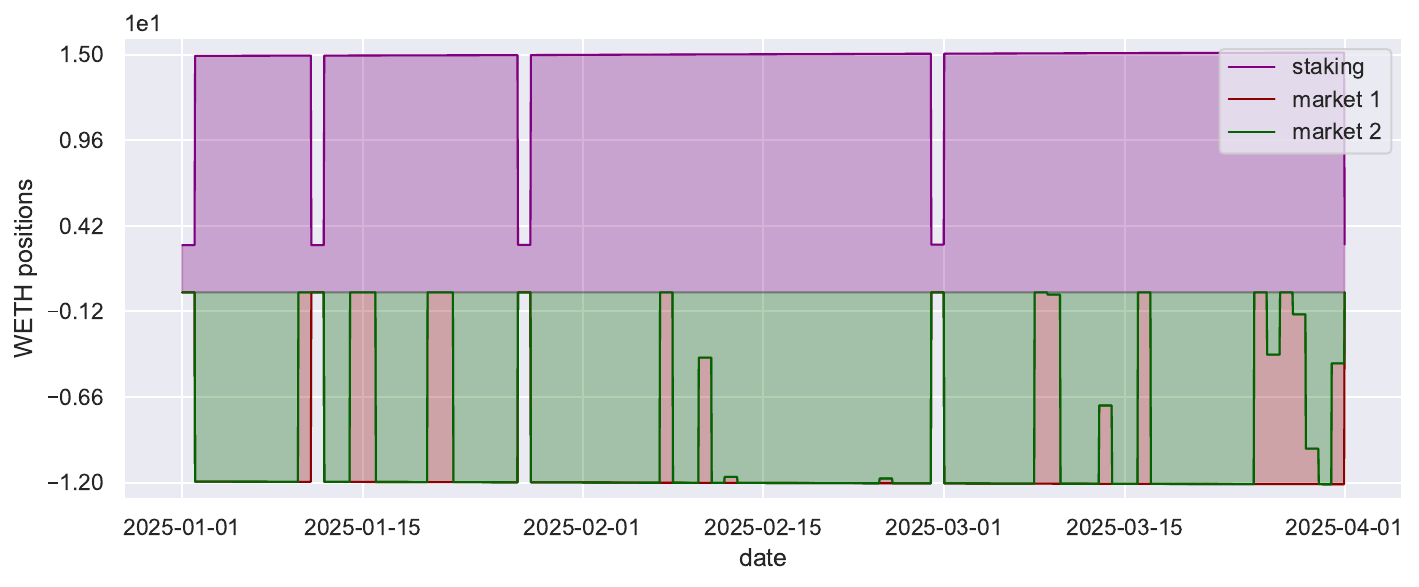}
        \caption{1d-freq rebalancing}
    \end{subfigure}
    \caption{Evolution of the WETH positions of the ``loopy'' (low cap) strategy on the Ethereum blockchain from January 1, 2025 to April 1, 2025.}
    \label{fig:backtesting_ethereum_low_cap}    
\end{figure}
Table \ref{tab:ethereum_strategies} reveals several key insights. First, the leveraged strategy significantly outperforms simple staking across all configurations, with APYs ranging from $3.7\%$ to $6.2\%$ compared to $3.1\%$ for staking alone, representing up to a twofold improvement. Second, capital size has a substantial impact on achievable returns: the low-cap strategy (\$10k) achieves nearly double the APY of the high-cap strategy (\$10m). This \emph{size effect} is a direct consequence of our theoretical framework: larger positions drive up pool utilization rates, which increases borrowing costs and reduces the spread between staking yield and borrowing rate. Third, rebalancing frequency matters more for smaller positions: hourly rebalancing improves APY by approximately $0.4$ percentage points for low-cap strategies ($6.2\%$ vs.\ $5.8\%$), while making virtually no difference for high-cap strategies ($3.7\%$ in both cases). This suggests that for large positions, the market impact of rebalancing dominates any benefits from faster rate arbitrage.
\begin{figure}
    \centering
    \begin{subfigure}{\linewidth} 
        \centering
        \includegraphics[scale = 0.6]{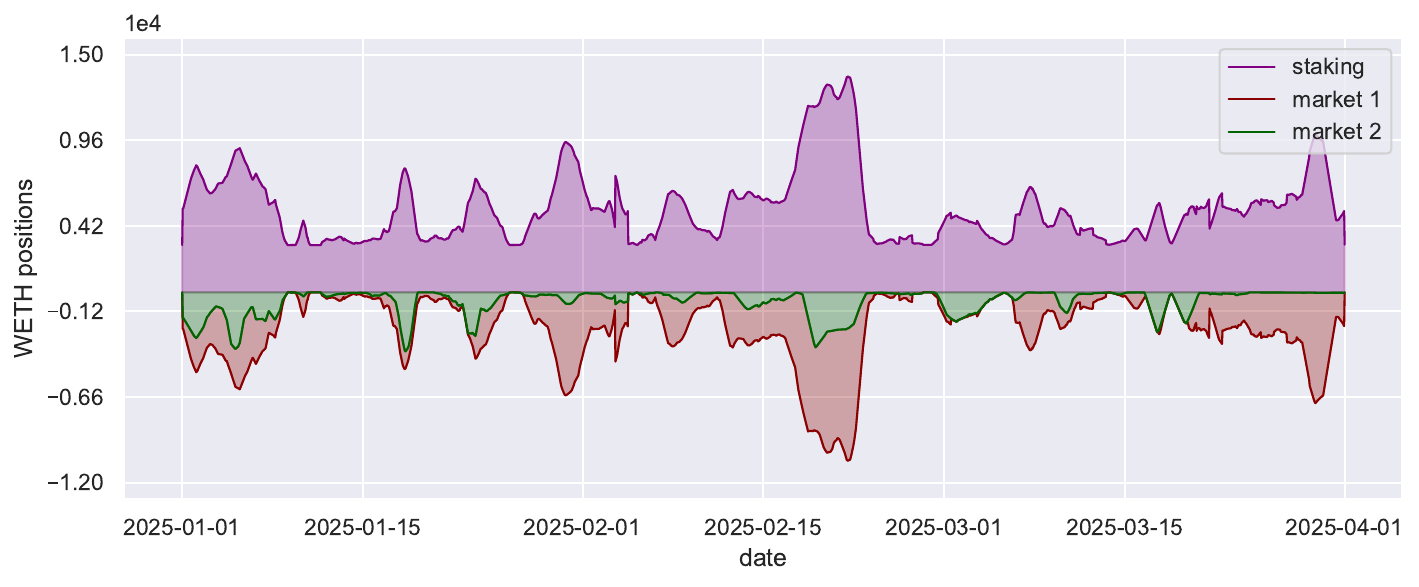}
        \caption{1h-freq rebalancing}
    \end{subfigure}
    \vfill
    \begin{subfigure}{\linewidth}
        \centering
        \includegraphics[scale = 0.6]{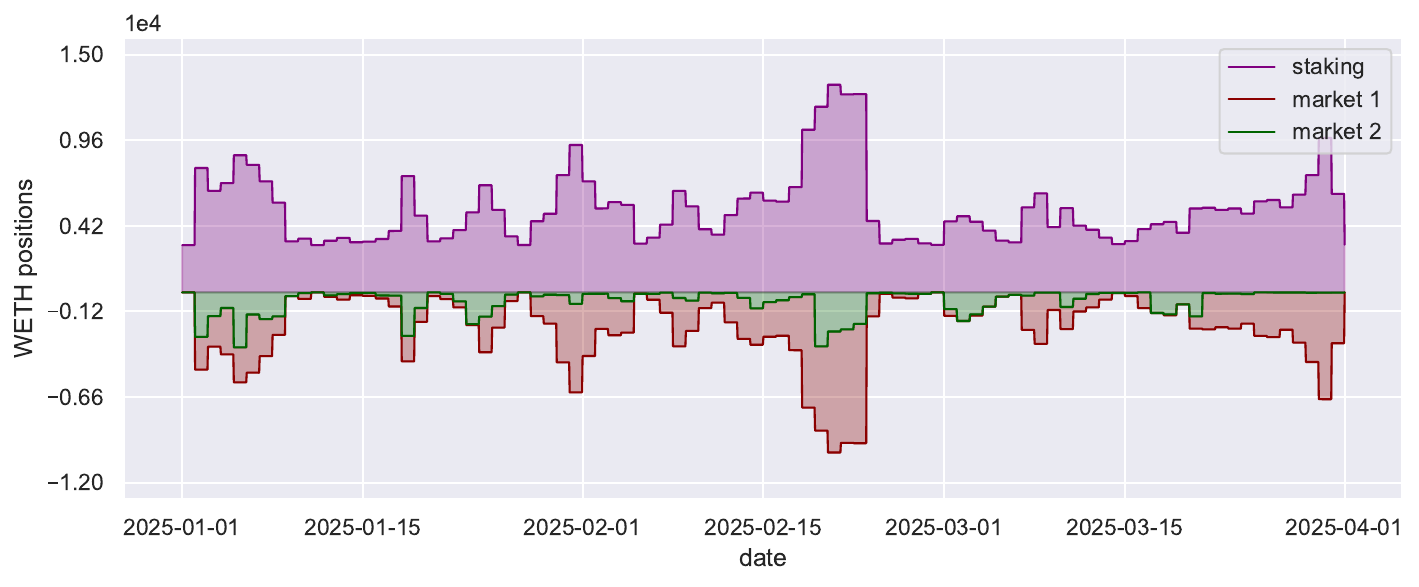}
        \caption{1d-freq rebalancing}
    \end{subfigure}
    \caption{Evolution of the WETH positions of the ``loopy'' (high cap) strategy on the Ethereum blockchain from January 1, 2025 to April 1, 2025.}
    \label{fig:backtesting_ethereum_high_cap}    
\end{figure}

Figures \ref{fig:backtesting_ethereum_low_cap}, \ref{fig:backtesting_ethereum_high_cap} illustrate the evolution of position allocations over the backtesting period, where negative positions correspond to the debt contracted on the markets. For low-cap strategies, the optimizer frequently reallocates between markets based on relative rate conditions, demonstrating active exploitation of rate differentials. In contrast, high-cap strategies show more stable allocations, as the size effect limits the profitability of aggressive rebalancing.

\begin{figure}
    \centering
    \includegraphics[scale = 0.6]{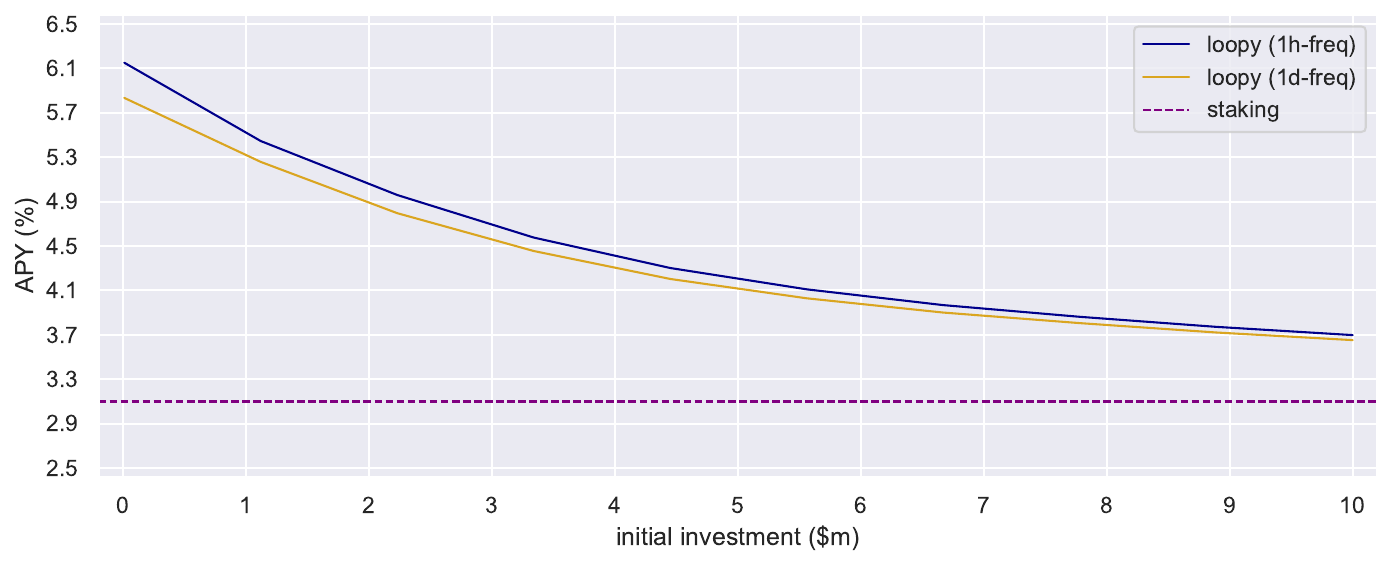}
    \caption{APY of the ``loopy'' strategy with respect to initial investment on the Ethereum blockchain from January 1, 2025 to April 1, 2025.}
    \label{fig:backtesting_ethereum_apys}
\end{figure}
Figure \ref{fig:backtesting_ethereum_apys} presents the APY as a function of initial investment, clearly illustrating the monotonically decreasing relationship predicted by our theoretical analysis. The curve exhibits a steep decline for investments below \$1m, after which returns asymptotically approach the simple staking rate as position size dominates pool utilization.
\begin{figure}
    \centering
    \includegraphics[scale = 0.6]{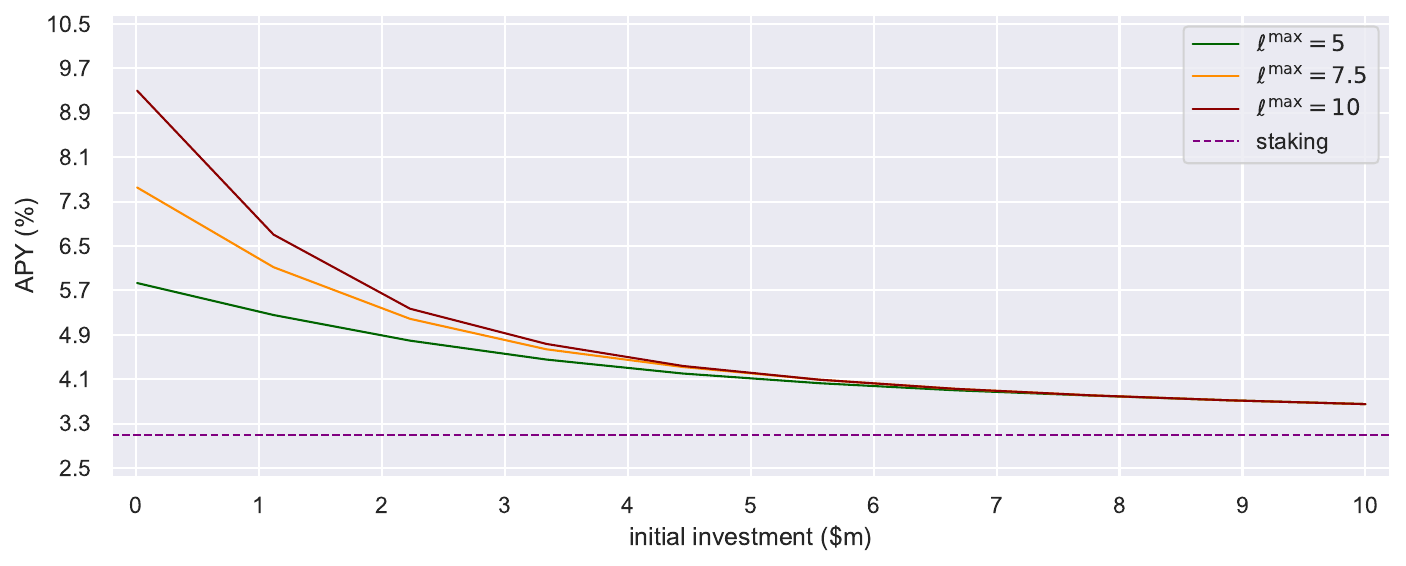}
    \caption{APY of the ``loopy'' (1d-freq) strategy with respect to initial investment on the Ethereum blockchain for different values of $\elm$ from January 1, 2025 to April 1, 2025.}
    \label{fig:backtesting_ethereum_lev_apys}
\end{figure}
Figure \ref{fig:backtesting_ethereum_lev_apys} presents the APY as a function of initial investment for different values of $\elm$ (1d-freq). While the differences across configurations are significant for small-cap investments, they become negligible for large-cap allocations, as market saturation prevents the strategy from exploiting the full investment capacity.

\begin{figure}
    \centering
    \includegraphics[scale = 0.6]{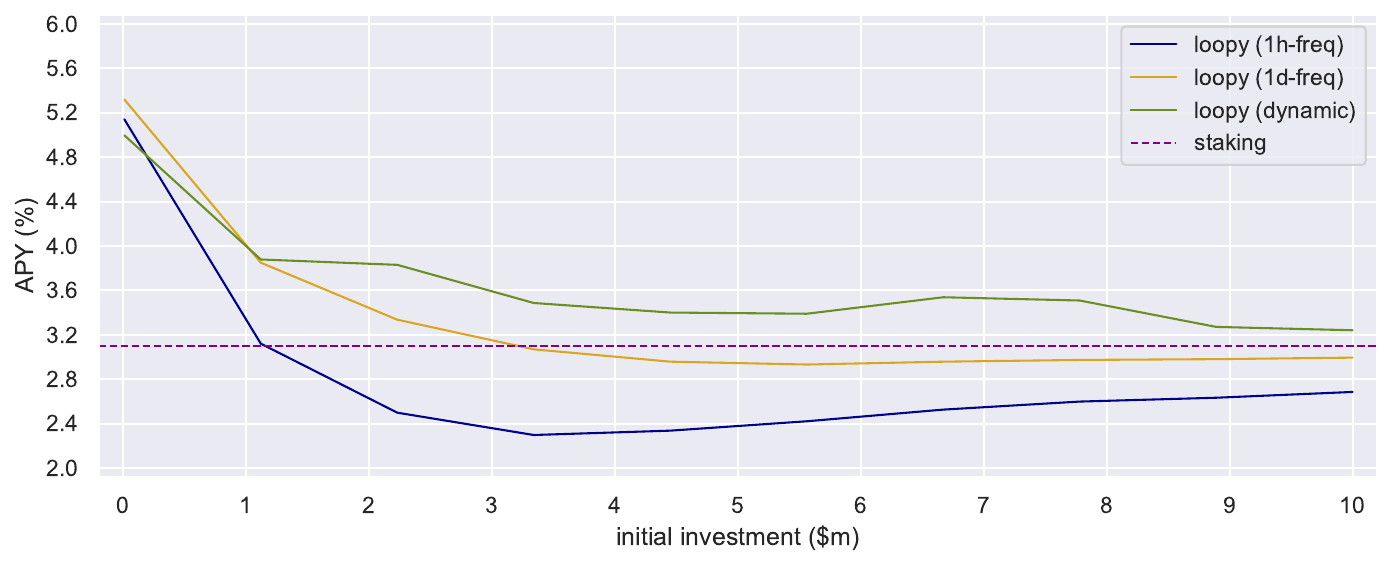}
    \caption{APY of the ``loopy'' strategy with respect to initial investment including fees on the Ethereum blockchain from January 1, 2025 to April 1, 2025.}
    \label{fig:backtesting_ethereum_apys_with_fees}
\end{figure}
Second, we introduce a $1$bp fee for selling wstETH for WETH. Both strategies—whether rebalanced daily or hourly—are significantly affected by the fee. Indeed, the high volatility of the borrowing rate can generate false signals, leading the strategy to rebalance, only to rebalance again shortly after. To address this issue, we introduce a third strategy, denoted ``loopy'' (dynamic), in which rebalancing occurs only if the expected yield after rebalancing exceeds the current yield by more than $20$bps. Figure \ref{fig:backtesting_ethereum_apys_with_fees} presents the APY as a function of initial investment. While loopy strategies—whether rebalanced daily or hourly—underperform beyond \$3m and \$1m invested, respectively, the dynamic strategy still outperforms native staking even with \$10m.

(See Appendix \ref{appendix:base} for the backtest on the Base blockchain.)

\section{Discussion and conclusion}\label{discussion}

This paper presents a comprehensive mathematical framework for optimizing leveraged staking strategies across multiple DeFi lending markets. We derived closed-form solutions for optimal allocation under three interest rate models (linear, kinked, and adaptive) and validated our theoretical results through backtesting on the Ethereum and Base blockchains. Our results demonstrate that optimal rebalancing can significantly enhance returns, with APYs reaching $6.2\%$ for small capital positions compared to $3.1\%$ for simple staking.

\paragraph{Key findings} Our backtesting results reveal several important insights. First, capital size has a substantial impact on achievable returns: smaller positions (\$10k) achieve significantly higher APYs than larger positions (\$10m) due to their reduced impact on pool utilization rates. Second, rebalancing frequency matters, though the marginal benefit diminishes beyond hourly rebalancing.

\paragraph{Limitations and assumptions} Our framework relies on several simplifying assumptions that warrant discussion. First, we assume complete liquidity; i.e., that positions can be opened, closed, and rebalanced instantaneously without slippage. While this is approximately true for small positions, large capital movements may face liquidity constraints in practice. Second, we treat the staking rate as exogenous and constant, whereas in reality it varies based on validator performance and network conditions. Third, and most importantly, our analysis assumes a single agent optimizing in isolation. When multiple sophisticated agents employ similar strategies, game-theoretic considerations become crucial.

\paragraph{Game-theoretic considerations} If many agents simultaneously pursue optimal leveraged strategies, a feedback loop emerges: increased borrowing demand raises utilization rates, which increases borrowing costs, which in turn reduces the profitability of leveraged positions. This creates a Nash equilibrium where individual optimization may lead to collectively suboptimal outcomes. Analyzing this multi-agent scenario requires extending our framework to incorporate strategic interactions, potentially using mean-field game theory or evolutionary game dynamics. This represents an important direction for future research.

\paragraph{Future directions} Several extensions of this work merit investigation. First, incorporating stochastic interest rate dynamics and staking rate volatility would enable risk-adjusted optimization and Value-at-Risk constraints. Second, analyzing the impact of liquidation risk under volatile collateral prices would provide more robust strategies for real-world deployment. Third, including multiple collateral types and cross-chain opportunities is a natural extension and would capture the full complexity of modern DeFi ecosystems. Finally, empirical analysis of how quickly capital actually moves in response to rate differentials would validate our complete liquidity assumption and inform practical rebalancing policies.

In conclusion, this work demonstrates that rigorous mathematical optimization can substantially improve DeFi investment strategies. The transparency and programmability of DeFi protocols enable a level of analytical precision impossible in traditional finance. As the DeFi ecosystem matures, we anticipate that such optimization frameworks will become standard tools for both individual investors and institutional participants.

%\section*{Acknowledgments}
%
%The authors would like to thank Romeo for fruitful discussions.
%
%\section*{Disclosure of interest}
%
%The authors have no competing interests to declare.

\bibliography{bibliographie.bib}

\section*{Appendix}

\appendix
\addtocontents{toc}{\protect\setcounter{tocdepth}{-5}}
\renewcommand*{\thesubsection}{\Alph{subsection}}

\subsection{Proof of equivalence between (\ref{eq:problem_x_l}) and (\ref{eq:problem_x_1_x_0})}\label{appendix:equivalence}

The action takes place in $\mathbb R^{2n}$ where $n$ is the number of markets at play. We pick a maximum leverage $\elm_{i} > 1$ for $i=1,\ldots,n$, and a budget size $\xi > 0$. The respective domains (thus parameterized) of the two problems are:
\begin{equation}
D_1(\xi,\elm) 
= 
\{
(x,\ell) \mid
x_{i} \ge 0,\,
\sum_{i=1}^{n} x_{i} = \xi,\,
\ell_{i} \ge 1,\,
\ell_{i} \le \elm_{i}
\}
\end{equation}
\begin{equation}
D_2(\xi) = 
\{
(x^0,x^1) \mid
x^0_i \ge 0,\,
x^1_i \ge 0,\,
\sum_{i=1}^{n} x^0_i + \sum_{i}^{n} x^1_i = \xi
\}
\end{equation}
We wish to show that the following map is a bijection between $D_1(\xi,\elm)$ and $D_2(\xi)$:
\begin{equation}
\theta(\elm):
(x_i,\ell_i) 
\mapsto
(
x_i \, \frac{\elm_i - \ell_i}{\elm_i - 1},
x_i \, \frac{\ell_i - 1}{\elm_i - 1}
)
\end{equation}
Since $\elm_i > 1$ this map is well-defined on $\mathbb R^{2n}$.

Let's first show that $\theta(D_1)\subseteq D_2$. Clearly, $x^0_i$, $x^1_i\ge0$. Also $x^0_i+x^1_i=x_i$, therefore the budget constraint is also satisfied.

We can invert $\theta$ on $\mathbb R^{2n}$:
\begin{equation}
\theta^{-1}:
(x^0_i,x^1_i) 
\mapsto
(
x^0_i+x^1_i,
\frac{x^0_i+\elm_i x^1_i}{x^0_i+x^1_i}
)
\end{equation}
and $\theta^{-1}$ is easily seen to restrict from $D_2$ to $D_1$.

To maximize the first objective $F_1$ on $D_1(\xi,\elm)$ is therefore equivalent to maximizing the second objective $F_2(\elm)$ on $D_2(\xi)$. Let $h_{i}(x) = xb_{i}(x)$, we have:
\begin{equation}
F_1(x,\ell) =
(\sum_{i=1}^{n} x_i\ell_i) s - \sum_{i=1}^{n} h_i(x_i(\ell_i-1)) 
\end{equation}
\begin{equation}
F_2(\elm)(x^0,x^1) =
(\sum_{i=1}^{n} x^0_i) s +
(\sum_{i=1}^{n} x^1_i\elm_i) s 
- \sum_{i=1}^{n} h_i(x^1_i(\elm_i-1)) 
\end{equation}
The claim above is true because $F_2(\elm)\circ\theta=F_1$ as can be readily verified:
\begin{equation}
F_2(\elm)(\theta(x,\ell)) =
s\sum_{i=1}^{n} x_i\ell_i - \sum_{i=1}^{n} h_i(x_i(\ell_i-1)) = F_1(x,\ell)
\end{equation}
Not only those problems have the same maximiser, but the max values are the same (they are the optimal cash flow).

\subsection{Proof of Proposition \ref{pr:linear_rate_solution}}\label{appendix:proof_linear_rate_solution}

Substituting the linear rate model \eqref{eq:linear_rate_model} into the first-order condition \eqref{eq:first_order_condition_i_unsaturated} yields: $x^{*}_{i}(\la) = \alpha_{i} \big [ \beta_{i} - \la \big ]^{+}$ for $i = 1, \ldots, n$, where the parameters are defined in \eqref{eq:parameters_linear_rate_solution}.

Let the markets be ordered according to \eqref{eq:beta_sorted}, and let $k \in \{ 1, \ldots, n \}$ be such that:
\begin{equation}\label{eq:condition_beta}
\beta_{k} \geq \la^{*} > \beta_{k+1}
\end{equation}
The optimal allocation then takes the form:
\begin{equation}\label{eq:temporary_linear_rate_solution}
x^{*}_{i}(\la^{*}) = \left\{
    \begin{array}{ll}
        \alpha_{i} \big [ \beta_{i} - \la^{*} \big ] & \mbox{if } i \leq k \\
        0 & \mbox{if } i > k
    \end{array}
\right.
\end{equation}
The budget constraint \eqref{eq:budget_constraint_unsaturated} becomes:
\begin{equation}
\sum^{n}_{j=1} x^{*}_{j}(\la^{*}) = \sum^{k}_{j=1} \alpha_{j} \big [ \beta_{j} - \la^{*} \big ] = \xi
\end{equation}
from which we deduce \eqref{eq:optimal_lambda}. Substituting \eqref{eq:optimal_lambda} into \eqref{eq:temporary_linear_rate_solution} yields the optimal allocation \eqref{eq:linear_rate_solution}. Finally, condition \eqref{eq:condition_beta} implies that:
\begin{equation}
\sum^{k}_{j=1} \alpha_{j} \big [ \beta_{j} - \beta_{k} \big ] < \xi \leq \sum^{k+1}_{j=1} \alpha_{j} \big [ \beta_{j} - \beta_{k+1} \big ]
\end{equation}
which is the condition \eqref{eq:phi_condition}.

\subsection{Proof of Proposition \ref{pr:kinked_rate_solution}}\label{appendix:proof_kinked_rate_solution}

We consider the maximization problem for a given market $i$ under the kinked rate model. The first-order optimality condition is derived from the KKT conditions, accounting for the non-differentiability of the objective function at the kink point. We examine the cases where the utilization rate is below and above the target rate separately.

\paragraph{Current utilization is below the target rate ($\bar{S}_{i} u^{*} - \bar{B}_{i} > 0$)} The kinked rate function \eqref{eq:kinked_rate_model} is piecewise linear with a discontinuity in its derivative at $x^{(\text{kink})}_{i} = \frac{\bar{S}_{i} u^{*} - \bar{B}_{i}}{\elm_{i} - 1}$. At this point, we must use the subdifferential of $b_{i}$ denoted $\partial b_{i}$. The subdifferential at $x^{(\text{kink})}_{i}$ is the interval between the left and right derivatives:
\begin{equation}
\partial b_{i} (x^{(\text{kink})}_{i}) = \Big [ \frac{r_{slope1}}{\bar{S}_{i} u^{*}}, \frac{r_{slope2}}{\bar{S}_{i} (1 - u^{*})} \Big ]
\end{equation}
The subdifferential optimality condition requires that there exists some $g \in \partial b_{i} (x^{(\text{kink})}_{i})$ such that:
\begin{equation}
\la = \elm_{i} s - ( \elm_{i} - 1 ) \big (b_{i}(x^{(\text{kink})}_{i} ( \elm_{i} - 1 )) + x^{(\text{kink})}_{i} ( \elm_{i} - 1 ) g \big )
\end{equation}
Substituting $b_{i}(x^{(\text{kink})}_{i}) = r_{base} + r_{slope1}$ and $x^{(\text{kink})}_{i} = \frac{\bar{S}_{i} u^{*} - \bar{B}_{i}}{\elm_{i} - 1}$ yields the condition:
\begin{equation}
\la \in \big [ \la^{2}_{i}, \la^{1}_{i} \big ]
\end{equation}
where $\la^{1}_{i}$ and $\la^{2}_{i}$ are defined in \eqref{eq:lambdas_kinked_rate_solution}.
Thus, when $\la \in \big [ \la^{2}_{i}, \la^{1}_{i} \big ]$, the optimal solution is $x^{(\text{kink})}_{i}$. For $\la$ outside this interval, the solution lies in the associated linear regime. The assumption $r_{slope1} < \frac{u^{*}}{1 - u^{*}} r_{slope2}$ ensures $\la^{2}_{i} < \la^{1}_{i}$, validating the ordering of the regimes.

\paragraph{Current utilization is at or above the target rate ($\bar{S}_{i} u^{*} - \bar{B}_{i} \leq 0$)} When current utilization is at or above the target, only the second linear region is relevant, reducing to a linear rate model framework.

\subsection{Proof of Corollary \ref{co:adaptive_rate_solution}}\label{appendix:proof_adaptive_rate_solution}

The adaptive rate model can be reformulated as the kinked rate model by assigning the parameters:
\begin{equation}\label{eq:mapping_adaptive_to_kinked}
r_{base} = \frac{r^{\mathrm{target}}_{t}}{k_{d}}, \quad r_{slope1} = r^{\mathrm{target}}_{t} \big ( 1 - \frac{1}{k_{d}} \big ), \quad r_{slope2} = r^{\mathrm{target}}_{t} \big ( k_{d} - 1 \big )
\end{equation}
Substituting \eqref{eq:mapping_adaptive_to_kinked} into Proposition \ref{pr:kinked_rate_solution} yields Corollary \ref{co:adaptive_rate_solution}.

\subsection{Backtest on the Base blockchain}\label{appendix:base}

\begin{figure}
    \centering
    \includegraphics[scale = 0.6]{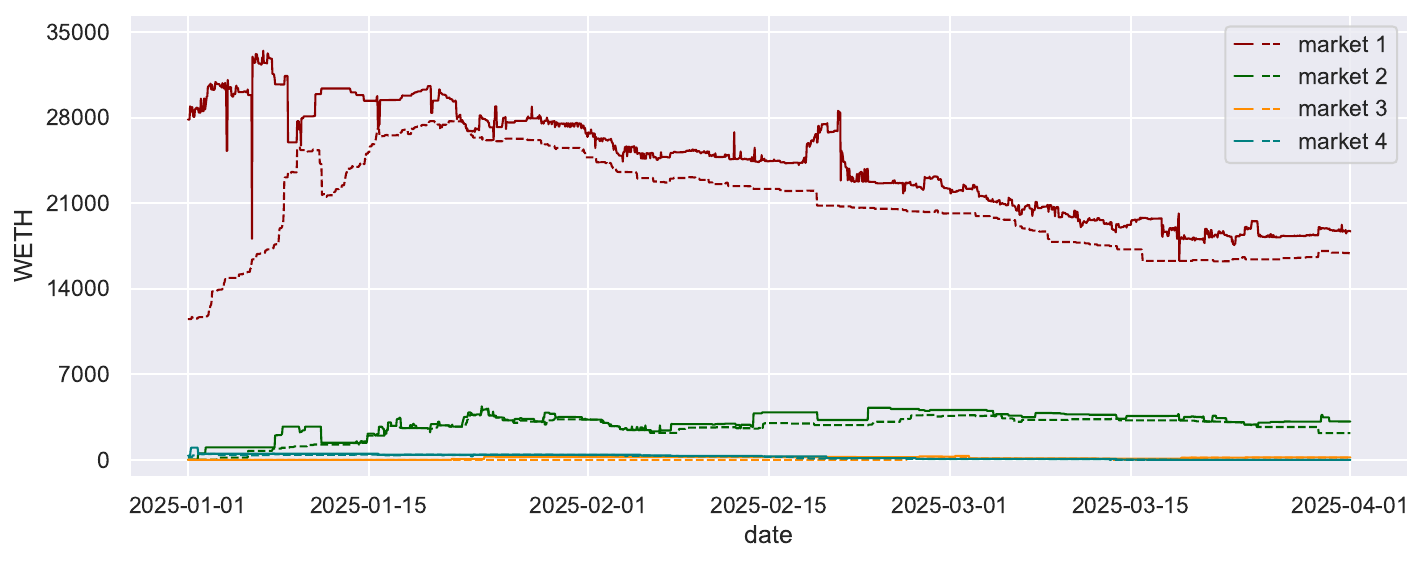}
    \caption{Evolution of WETH reserves (solid line: supplied funds; dashed line: borrowed fund) for the four largest wstETH/WETH markets on Morpho on the Base blockchain from January 1, 2025 to April 1, 2025.}
    \label{fig:evolution_base_market_reserves}
\end{figure}

\begin{table}
\centering
\begin{tabular}{cccc}
\toprule
market & ID & creation date & LLTV ($\%$) \\
\midrule
1 & \texttt{b991f6fd-568b-4332-998e-3fedf6afae20} & June 11, 2024 & $94.5$ \\
2 & \texttt{130cec4d-4fe4-4fbb-9d85-5e2c279eb854} & May 30, 2024 & $96.5$ \\
3 & \texttt{13aac762-a267-4d6e-8904-9f886babec7f} & July 22, 2024 & $94.5$ \\
4 & \texttt{58e6612e-a221-4b64-b53c-6b69c3a3836e} & May 30, 2024 & $94.5$ \\
\bottomrule
\end{tabular}
\caption{IDs, creation dates and LLTVs of the four largest wstETH/WETH markets on Morpho on the Base blockchain.}
\label{tab:base_markets}
\end{table}
On Base, we select the four largest Morpho markets for the wstETH/WETH pair over the period from January 1, 2025, to April 1, 2025. Although the last two are negligible compared to the top two (see Figure \ref{fig:evolution_base_market_reserves}), we retain them to illustrate a backtest involving more markets than in the Ethereum case. The characteristics of each market are summarized in Table \ref{tab:base_markets}.

\begin{figure}
    \centering
    \begin{subfigure}{\linewidth} 
        \centering
        \includegraphics[scale = 0.6]{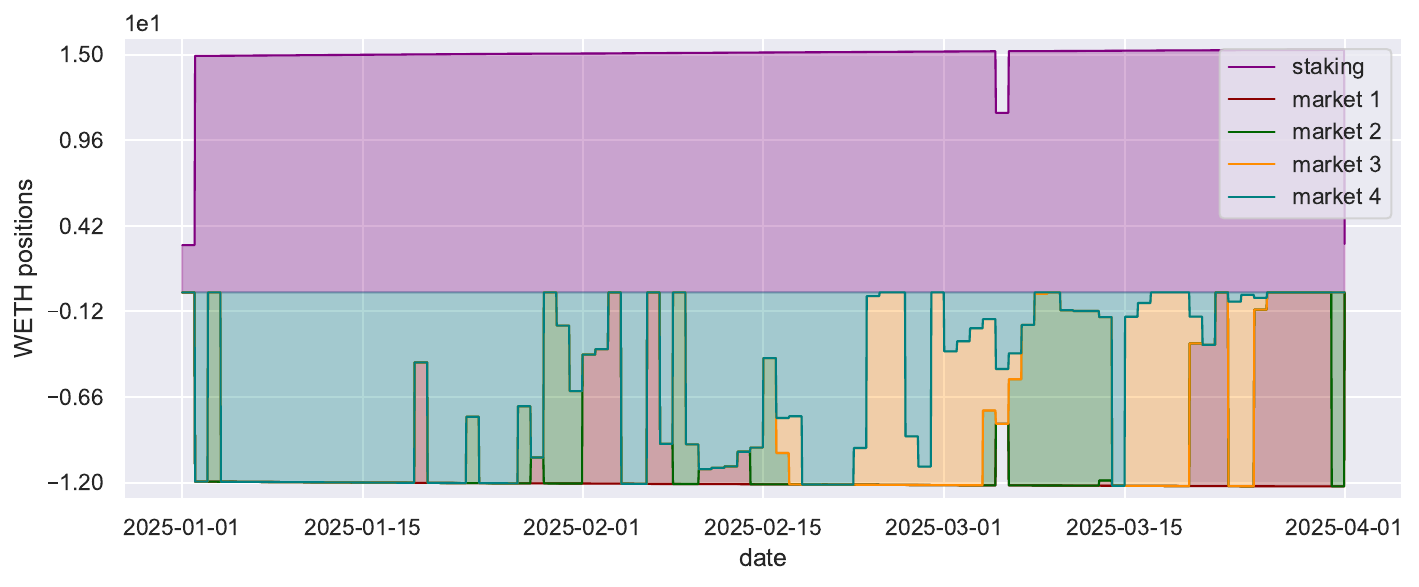}
        \caption{low cap}
    \end{subfigure}
    \vfill
    \begin{subfigure}{\linewidth}
        \centering
        \includegraphics[scale = 0.6]{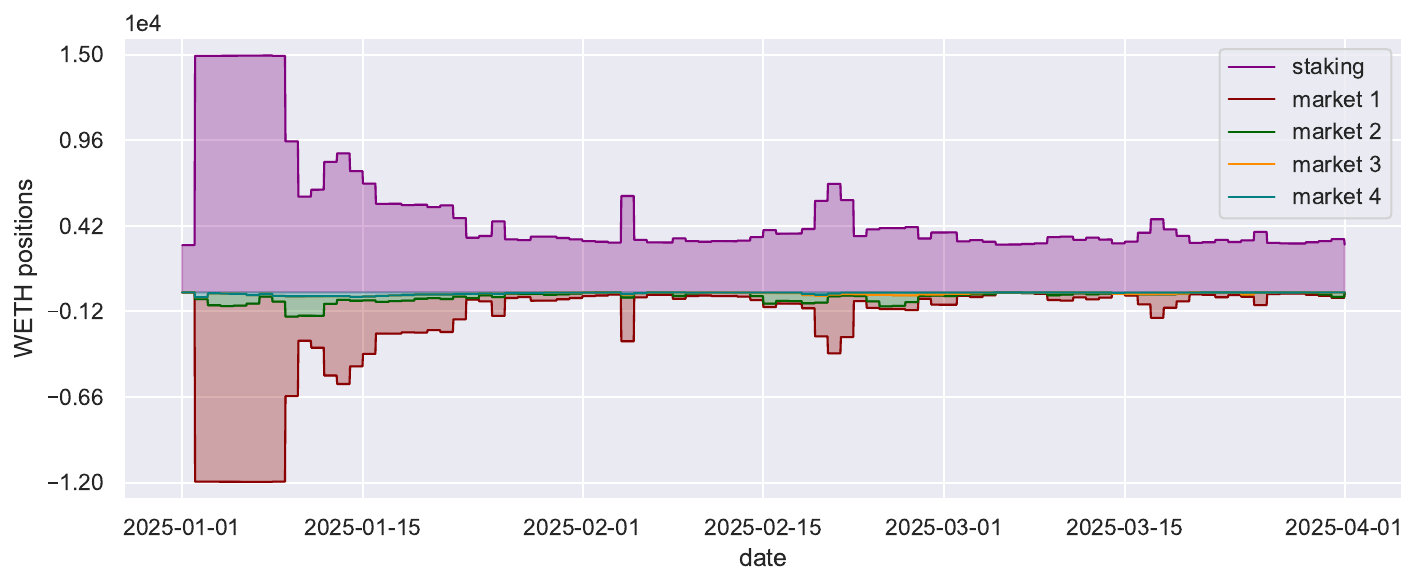}
        \caption{high cap}
    \end{subfigure}
    \caption{Evolution of the WETH positions of the ``loopy'' (1d-freq) strategy on the Base blockchain from January 1, 2025 to April 1, 2025.}
    \label{fig:backtesting_base}    
\end{figure}
Figure \ref{fig:backtesting_base} illustrates the evolution of position allocations over the backtesting period for the low- and high-cap 1d-frequency strategies. The low-cap strategy loops through all four markets, whereas the high-cap one almost exclusively loops through the first market, as the others quickly become saturated.

\begin{figure}
    \centering
    \includegraphics[scale = 0.6]{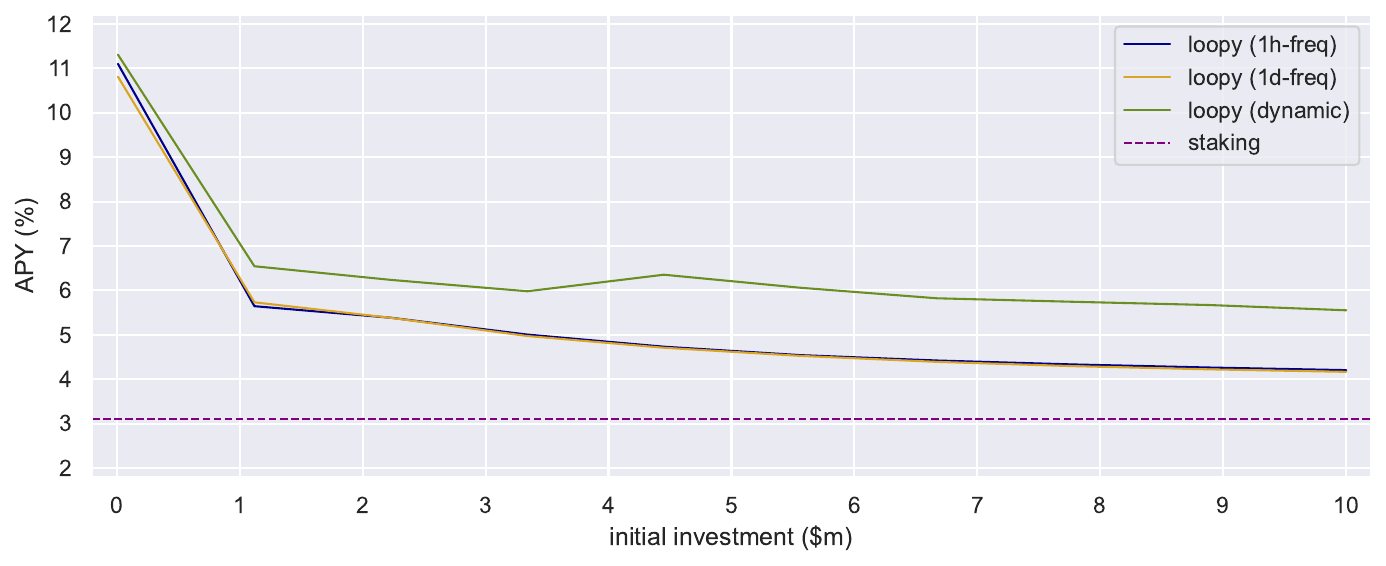}
    \caption{APY of the ``loopy'' strategy with respect to initial investment including fees on the Base blockchain from January 1, 2025 to April 1, 2025.}
    \label{fig:backtesting_base_apys_with_fees}
\end{figure}
Figure \ref{fig:backtesting_base_apys_with_fees} presents the APY as a function of the initial investment including a $1$bp fee for selling wstETH for WETH in the backtest. As in the Ethereum case, the dynamic strategy outperforms both daily and hourly rebalancing strategies.

\end{document}